\documentstyle[12pt,epsf]{article}
\advance\textheight by \topskip
\begin{document}
\baselineskip=0.45cm
\thispagestyle{empty}
\begin{flushright}
KUCP0187\\
August 1, 2001\\
\end{flushright}
\vskip 2 cm
\begin{center}
{\LARGE\bf Bulk Gravitational Field and  \\ 
Cosmological Perturbations on the Brane}
\vskip 1.7cm

 {\bf Kazuya Koyama}\footnote{
E-mail: kazuya@phys.h.kyoto-u.ac.jp} 
{\bf Jiro Soda}
\footnote{E-mail: jiro@phys.h.kyoto-u.ac.jp } \

\vskip 1.5mm

\vskip 2cm
 $^1$ Graduate School of Human and Environment Studies, Kyoto University, 
       Kyoto  606-8501, Japan \\
 $^2$ Department of Fundamental Sciences, FIHS, Kyoto University,
       Kyoto, 606-8501, Japan \\
\end{center}

\def\k{\mbox{\boldmath$k$}}
\def\x{\mbox{\boldmath$x$}}

{\centerline{\large\bf Abstract}}
\begin{quotation}
\vskip -0.4cm
We investigate the effect of the bulk gravitational field on the
cosmological perturbations on a brane embedded in the 
5D Anti-de Sitter (AdS) 
spacetime. The effective 4D Einstein equations 
for the scalar cosmological perturbations on the brane are obtained
by solving the perturbations in the bulk. Then the behaviour of the 
corrections induced by the bulk gravitational field 
to the conventional 4D Einstein equation are 
determined. Two types of the corrections are found. 
First we investigate the corrections which  
become significant at scales below the AdS curvature scales
and in the high energy universe with the energy density 
larger than the tension of the brane.
The evolution equation for the perturbations on the brane is 
found and solved. 
Another type of the corrections is induced on the brane 
if we consider the bulk perturbations which do not contribute to the
metric perturbations but do contribute to the matter 
perturbations. At low energies, they have imaginary 
mass $m^2=-(2/3) \k^2$ in the bulk where $\k$ is the 3D
comoving wave number of the perturbations. They diverge
at the horizon of the AdS spacetime.  
The induced density perturbations behave as sound waves with sound 
velocity $1/\sqrt{3}$ in the low energy universe. 
At large scales, they are 
homogeneous perturbations that depend only on time and decay like 
radiation. They can be identified as the perturbations of the dark 
radiation. They produce isocurvature perturbations in the matter dominated 
era. Their effects can be observed as the shifts of the location and
the height of the acoustic peak in the CMB spectrum.

\end{quotation}
 \newpage

\def\k{\mbox{\boldmath$k$}}
\def\x{\mbox{\boldmath$x$}}

\baselineskip=0.56cm

\section{Introduction}
\setcounter{equation}0
\hspace{0.5cm}
Recent developments of the particle physics revive the
old idea that we are living in 4D brane in higher dimensional
spacetime \cite{BW1,BW2}. 
Since Randall and Sundrum proposed fascinating
model for the brane world, many works have been done about 
the consistency of the model with observations \cite{RS}. 
In their model, our 3-brane universe is located in the 
5D Anti-de Sitter (AdS) spacetime. 
The essence of the model is that the spacetime is effectively
compactified with curvature scale $l$ of the AdS spacetime. 
Thus even the gravity can propagate in the whole
higher dimensional spacetime, the 4D Newtonian gravity is 
reproduced at the scales larger than $l$ on the brane. 

After their work, the cosmological consequences of the model 
are actively 
investigated [4-15]. The set up of the model is given as follows.
The action describing the brane world picture is given by
\begin{equation}
S= \frac{1}{2 \kappa^2}\int d^5 x \sqrt{-g}
\left(
{\cal R}^5 +  \frac{12}{l^2} \right)
- \sigma \int d^4 x \sqrt{-g_{brane}}
+ \int d^4 x \sqrt{-g_{brane}} {\cal L}_{matter},
\label{0-1}
\end{equation}
where ${\cal R}^5$ is the 5D Ricci scalar, $l$ is the 
curvature radius of the AdS spacetime and $\kappa^2=8 \pi G_5$ 
where $G_5$ is the Newton's constant in the 5D spacetime. 
The brane has tension $\sigma$ and the induced metric on the brane 
is denoted as $g_{brane}$. The tension $\sigma$ of the brane is 
taken as $\kappa^2 \sigma=6/l$ to ensure that the brane becomes
Minkowski spacetime if there is no matter on the brane. 
Matter is confined to the 4D brane world and is described by the
Lagrangian ${\cal L}_{matter}$. We will assume
$Z_2$ symmetry across the brane. 
It has been shown that the spatially homogeneous and isotropic
universe can also be embedded in this model. In order to
study further consistency of the model with the observations, 
it is needed to study the behaviour of the cosmological 
perturbations [16-18]. 
The cosmological perturbations in the brane world provide 
useful tests for the brane world idea. This is because the 
perturbations in the brane world interact with the bulk gravitational
field, which is the inherent nature of perturbations in the brane world.
Several formalisms and applications have been
developed [19-29]. Especially, we showed that 
the evolution of the perturbations is the same as the one obtained
in the conventional 4D theory 
at low energies when the Hubble horizon of the brane universe 
is larger than $l$. We also pointed out that at high energies,
the evolution of the perturbations changes significantly 
\cite{KJ}. 

The purpose of this paper is to make further clarification
about the difference between the behaviour of the perturbations
in the brane world model and the one in the conventional 4D model. 
For this purpose, it is desirable to obtain the effective 4D Einstein
equations on the brane. 
There are several works which investigate the effective 4D Einstein 
equations using the covariant method [30,31].  
The effective 4D Einstein equations are obtained as 
\begin{equation}
G_{\mu \nu} + {\cal E}_{\mu \nu}
=\frac{\kappa^2}{l} T_{\mu \nu} + \kappa^4 \Pi_{\mu \nu},
\label{1-1-1}
\end{equation}
where
\begin{equation}
\Pi_{\mu \nu}=-\frac{1}{4}T_{\mu \alpha}T_{\nu}^{\alpha}
+\frac{1}{12} T^{\alpha}_{\alpha}T_{\mu \nu}+
\frac{1}{24}(3 T_{\alpha \beta}T^{\alpha \beta}- 
(T^{\alpha}_{\alpha})^2)
g_{\mu \nu},
\end{equation}
and ${\cal E}_{\mu \nu}$ is the projected 5D Weyl tensor. 
In \cite{Large}, the large scale cosmological perturbations 
are analyzed with (\ref{1-1-1}). They study the evolution of 
perturbations using equations solely on the brane and without 
solving the perturbations in the bulk. Although 
significant results are obtained, their approach is clearly
limited because the behaviour of ${\cal E}_{\mu \nu}$ cannot
be determined without solving the bulk perturbations.  
In the previous paper \cite{KJ}, we have developed a method to solve the
perturbations in the bulk. In this paper, using the method, 
we obtain the effective Einstein equation
for the scalar cosmological perturbations by solving the 
perturbations 
in the bulk (\ref{3-6}) and (\ref{3-7}). 
Then we can determine the behaviour of the 
corrections induced by the bulk gravitational field 
to the conventional 4D Einstein equations . This is an essential 
part of the work predicting the CMB anisotropies in the brane world.

We will obtain the effective Einstein equations in two ways.
First we derive the effective Einstein equations from the equations 
solely on the brane as in \cite{Large}. We observe the limitations
of this method. Then we construct the effective Einstein equations 
again by solving the bulk perturbations. 
The evolution of the perturbations on the brane is investigated
using the effective Einstein equations. 
We concentrate our attention on the scalar 
perturbations on a brane in the AdS spacetime. 
A new type of corrections arises
if we choose appropriate boundary conditions on the perturbations in 
the bulk so that the perturbations do not contribute to the
metric perturbations but do contribute to the matter perturbations.
They induce the density perturbations on the brane
which behave as sound waves with sound 
velocity $1/\sqrt{3}$ in the low energy universe.
At large scales, they are 
homogeneous perturbations that depend only on time and decay like 
radiation. 
We will discuss the effects of these perturbations
on the CMB spectrum. 

The structure of the paper is as follows.
In section 2, we construct the effective Einstein equation 
for the background spacetime in two ways as an example.
In section 3, the effective Einstein equations for perturbations 
are constructed from the equations on the brane. 
Two types of the corrections to the conventional 4D Einstein 
equations are found. We see a complete set of the effective 4D 
Einstein equations cannot be derived from equations solely on
the brane.
In section 4, the effective Einstein equations are obtained again 
by solving the perturbations in the bulk and imposing the junction 
conditions. A complete set of the equations is obtained. 
We find again two types of the corrections, but now 
they are obtained according to the boundary conditions of the 
perturbations in the bulk.
In section 5, we take the boundary
condition that the perturbations do not diverge at the horizon 
of the AdS spacetime. And we investigate the modifications of the 
evolution. In section 6, we allow the existence of the
perturbations which do not contribute to the metric 
perturbations but do contribute to 
the matter perturbations. 
The modifications of the evolution caused by
these perturbations are studied. In section 7, 
we summarize the results. In Appendix A, the equations used 
in section 2 are derived. In Appendix B, we review the formalism 
to solve the perturbations in the
bulk and impose the junction conditions. Then the effective Einstein 
equations are obtained. In Appendix C, the generation of the 
primordial fluctuations is discussed. The Mukhanov equation 
for the inflaton confined to the brane is obtained.

\section{Background spacetime}
\setcounter{equation}0
\hspace{0.5cm} It would be instructive to consider the background 
spacetime as an example for constructing the effective Einstein 
equations. We take the background metric as
\begin{equation}
ds^2=e^{2 \beta(y,t)}(dy^2-dt^2)+ e^{2 \alpha(y,t)}
\delta_{ij} dx^i dx^j.
\label{1-1}
\end{equation}
We will denote the power series expansion near the brane 
as 
\begin{equation}
\alpha(y,t)=\alpha_0(t)+ \alpha_1(t) \vert y \vert  + 
\frac{\alpha_2(t)}{2} y^2+ 
\cdot \cdot \cdot.
\label{1-2}
\end{equation}
The tension $\sigma$ of the brane is taken as $\kappa^2 \sigma=6/l$
and the 5D energy momentum tensor of the matter confined to the
brane is taken as
\begin{equation}
T^M_N= diag(0,-\rho,p,p,p) \delta(y). 
\label{1-3}
\end{equation}
The calculations
which are necessary to obtain the equations used in the
following discussions are performed in Appendix A. 

The first method is to use the power series expansion of the
5D Einstein equation. From the junction conditions, the first
derivatives of the metric perturbations with respect to
$y$ are written by the 
matter perturbations on the brane. Then we can obtain 
the equations about the variables on the brane
from the 5D Einstein equations that do not contain the second 
derivatives of the metric perturbations with respect to $y$. 
From these equations on the brane, 
we can construct the effective Einstein
equations on the brane. The junction conditions are given by
\begin{eqnarray}
\alpha_1(t) &=& - \frac{1}{l}   
- \frac{\kappa^2 \rho(t)}{6}, \nonumber\\
\beta_1(t) &=&  - \frac{1}{l}+ 
\frac{\kappa^2 \rho(t)}{3} + \frac{ \kappa^2 p(t)}{2},
\label{1-4}
\end{eqnarray}
where we take $e^{\beta_0(t)}=1$. The equations for
$\alpha_0$ and $\rho$ can be obtained from the power
series expansion of the 5D Einstein equation near the brane.
The $y^0$-th order of the $(y,0)$ and $(y,y)$ components 
are given by
\begin{eqnarray}
&& \ddot{\alpha}_0+2 \dot{\alpha}_0^2= \frac{\kappa^2}{2 l}
\left(\frac{\rho}{3}-p \right) - \frac{\kappa^4 \rho(\rho+3 p)}{36},
\nonumber\\
&& \dot{\rho}+ 3 \dot{\alpha}_0 (\rho+p)=0.
\label{1-5}
\end{eqnarray}
At low energies $\rho/\sigma \sim \kappa^{-2} l \rho \ll 1$, 
the former is identical with the trace part of the conventional 
4D Einstein equations with
\begin{equation}
8 \pi G_4=\kappa^2/l,
\end{equation}
where $G_4$ is the Newton's constant in the 4D spacetime.
The latter is the usual energy-momentum conservation of the
matter. The integration of the first equation gives the effective
Friedmann equation;
\begin{equation}
\dot{\alpha}_0^2 = \frac{\kappa^2}{3 l} \rho + 
\frac{\kappa^4 \rho^2}{36} + e^{-4 \alpha_0} C_0 ,
\label{1-6}
\end{equation}
where $C_0$ is the constant of the integration. 
This is the $(t,t)$ component of the effective Einstein
equations. In the 4D Einstein theory,  
$(t,t)$ component of the Einstein equations gives
\begin{equation}
\dot{\alpha}_0^2=\frac{8 \pi G_4}{3} \rho.
\label{1-7}
\end{equation}
Thus the constant of the integration $C_0$ should have been  
$0$ in order to match to the 4D Einstein theory at low energies. 
However the non-zero $C_0$ is not forbidden in the brane world.
Indeed, it is known that $C_0$ is related to the mass of the 5D 
AdS-Schwartzshild Black Hole. 
Thus the non-zero $C_0$ indicates the effect of 
the bulk.
The lesson is that even if we have a complete set of the equations 
for $\alpha_0$ and $\rho$ (\ref{1-5}) which are identical with
those in the conventional 4D theory at low energies, 
the correction to the Friedmann equation can exist. 
Because the term proportional to $C_0$ in (\ref{1-6}) 
behaves like radiation, it is often called dark radiation. 
The important point is that we cannot determine $C_0$ 
from the equations on the brane (\ref{1-5}). 
We need a different method to determine the correction 
which describes the effect of the bulk. 

Another way is to solve the 5D Einstein equation in the bulk. 
We should impose the boundary conditions (\ref{1-4}) on the brane.
The equations for $\beta$ and $\alpha$ in the 
bulk are given by
\begin{eqnarray}
-\ddot{\beta}+\beta''-\frac{1}{l^2} e^{2 \beta}&=& 0, \nonumber\\
-\ddot{\alpha}+\alpha''-\frac{1}{l^2} e^{2 \beta} &=& 0,
\label{1-8}
\end{eqnarray}
where we assumed the bulk is purely 
AdS spacetime without Schwartzshild mass (cf. Appendix A).
We can obtain the solution 
\begin{equation}
e^{2 \beta(y,t)} =  4 \frac{f'(u)g'(v)}{(f(u)-g(v))^2}, \quad
e^{2 \alpha(y,t)} = \frac{1}{(f(u)-g(v))^2},
\label{1-9}
\end{equation}
where $u=(t-y)/l,v=(t+y)/l$ and $f(u)$ and $g(v)$ are the 
arbitrary functions. Thus
the matter on the brane is written by $f$ and $g$
from the junction conditions (\ref{1-4}) as
\begin{equation}
\alpha_1 =\frac{1}{l} \left( \frac{f(t/l)'+g(t/l)'}{f(t/l)-g(t/l)} 
\right)
=-\frac{1}{l}-
\frac{\kappa^2 \rho}{6}.
\label{1-10}
\end{equation}
In usual, we find the solutions of $f$ and $g$ from the junction condition 
(\ref{1-10}). However, it is difficult to find the solutions
for perturbations in this way. Thus we propose a new way to 
find the solutions, namely transforming
the junction condition (\ref{1-10}) to the
effective Einstein equation.
From (\ref{1-9}), we obtain  
\begin{eqnarray}
\dot{\alpha}_0 &=&-\frac{1}{l} 
\left( \frac{f(t/l)'-g(t/l)'}{f(t/l)-g(t/l)} \right), \nonumber\\
e^{2 \beta_0} &=& 4 \frac{f(t/l)' g(t/l)'} {(f(t/l)-g(t/l))^2}=1.
\label{1-11}
\end{eqnarray}
Then the term written by $f$ and $g$ in (\ref{1-10}) can be written
by the metric $\alpha_0$. 
We find $\alpha_1^2=\dot{\alpha}_0^2+1/l^2$. Thus the junction
condition (\ref{1-10})
gives the effective Friedmann equation on the brane
\begin{equation}
\dot{\alpha_0}^2=\frac{\kappa^2}{3 l}\rho+\frac{\kappa^4 \rho^2}{36}.
\label{1-12}
\end{equation}
Comparing this with (\ref{1-6}), 
we can determine $C_0=0$ for AdS background. 
We will consider the perturbations in this background.


\section{Effective Einstein equations from equations on the \\ brane}
\setcounter{equation}0
\hspace{0.5cm}
In this section we will derive the effective Einstein equations
for scalar cosmological perturbations using the equations on the brane.
The perturbed 5D energy-momentum tensor is taken as
\begin{equation}
\delta T^M_N = 
\left(
\begin{array}{ccc}
0 & 0 & 0 \\
0 & -\delta \rho & -(\rho+p)e^{\alpha_0} v_{,i} \\
0 & (\rho+p) e^{- \alpha_0} v_{,i}  & \delta p \: \delta_{ij}\\
\end{array}
\right) \: \delta (y),
\label{2-1}
\end{equation}
where we assume the isotropic stress of the matter perturbations
vanishes. The perturbed metric on the brane is taken as
\begin{equation}
ds_{brane}=-(1+2 \Phi_0) dt^2+e^{2\alpha_0(t)}
(1-2 \Psi_0) \delta_{ij} dx^i dx^j.
\label{2-2}
\end{equation}
The equations on the brane are obtained from $(y,y)$, $(y,0)$ 
and $ (y,i)$
components of the 5D Einstein equations as \cite{KJ}
\begin{eqnarray}
\ddot{\Psi}_0+4 \dot{\alpha}_0 \dot{\Psi}_0 +\dot{\alpha}_0 \dot{\Phi}_0
+2(\ddot{\alpha}_0+2 \dot{\alpha}_0^2) \Phi_0 
- \frac{1}{3} e^{-2 \alpha_0} (2 \nabla^2 \Psi_0 -\nabla^2 \Phi_0)
\nonumber\\
=
\frac{\kappa^2}{3} \left(\frac{\beta_1}{2} 
\delta \rho -\frac{3 \alpha_1}{2} \delta p \right),
\label{2-3}
\end{eqnarray}
\begin{equation}
\dot{\delta \rho} =(\rho+p)(3 \dot{\Psi}_0+ e^{-\alpha_0} 
\nabla^2 v)
-3 \dot{\alpha}_0 
\left( \delta \rho+ \delta p \right),
\label{2-4}
\end{equation}
\begin{equation}
\left( (\rho+p) e^{\alpha_0} v\right)^{\cdot}
= -3 \dot{\alpha}_0 (\rho+p) e^{\alpha_0} v + \delta p 
+(\rho+p) \Phi_0.
\label{2-5}
\end{equation}
As in the background, (\ref{2-3}) is the same as the 
trace of the conventional 4D Einstein equations at low energies
and (\ref{2-4}) and (\ref{2-5}) are the usual 
energy-momentum conservation for
the matter perturbations.
From these equations, we construct the effective 
4D Einstein equations. The Einstein equation gives a relation
between the matter perturbations and the metric perturbations.
Thus we try to write the matter perturbations 
in terms of the metric perturbations from (\ref{2-3}), (\ref{2-4}), 
and (\ref{2-5}). The equations can be regarded as the differential
equations for $\delta \rho$, $\delta p$ and $v$ with the source 
given by $\Phi_0$ and $\Psi_0$. Then the solutions for 
$\delta \rho$, $\delta p$ and $v$ are given by special solutions 
written by $\Phi_0$ and $\Psi_0$ and homogeneous solutions which
are independent of $\Phi_0$ and $\Psi_0$. The homogeneous 
solutions satisfy
\begin{eqnarray}
\dot{\delta \rho} &=& (\rho+p)e^{-\alpha_0} 
\nabla^2 v
-3 \dot{\alpha}_0 
\left( \delta \rho+ \delta p \right), \nonumber\\
\delta p &=&
\frac{1}{3}\left(1+\frac{\ddot{\alpha}_0}{\alpha_1^2} \right)
\delta \rho, \nonumber\\
\left( (\rho+p) e^{\alpha_0} v\right)^{\cdot}
&=&  -3 \dot{\alpha}_0 (\rho+p) e^{\alpha_0} v + \delta p,
\label{2-6}
\end{eqnarray}
where we used the background equations (\ref{A-12}).
From these equations, we can construct the second order
differential equation for $\delta \rho$.
Putting
\begin{equation}
\delta \rho= -\frac{1}{\alpha_1 l} e^{-4 \alpha_0} \chi,
\label{2-7}
\end{equation}
we obtain the equation for $\chi$ as 
\begin{equation}
\ddot{\chi}+ \dot{\alpha}_0 \left(1 -
\frac{\ddot{\alpha}_0}{\alpha_1^2} \right) \dot{\chi}-
\frac{1}{3}\left(1+\frac{\ddot{\alpha}_0}{\alpha_1^2}
\right) e^{- 2 \alpha_0} \nabla^2 \chi =0.
\label{2-8}
\end{equation}
The special solutions can be obtained perturbatively
assuming $\vert e^{-2\alpha_0} \nabla^2 \Psi /\ddot{\Psi}_0 \vert
\ll 1$. 
The solutions up to the order $\nabla^4 \Psi_0$ including the homogeneous
solutions written by $\chi$ are given by 
\begin{eqnarray}
-\frac{\kappa^2 \alpha_1}{2} \delta \rho
&=& -3 (\dot{\alpha}_0 \dot{\Psi}_0 + \dot{\alpha}_0^2 \Phi_0)
+e^{-2 \alpha_0} \nabla^2 \Psi_0 
-\frac{\kappa^2 \alpha_1}{2} \delta \rho^{(4)}
+ \frac{\kappa^2}{2 l}
\delta \rho_{\chi},
\nonumber\\
-\frac{\kappa^2 \alpha_1}{2} \delta p
&=& \ddot{\Psi}_0+\left(3 \dot{\alpha}_0-\frac{\dot{\alpha_0} 
\ddot{\alpha}_0}{\alpha_1^2}\right) \dot{\Psi}_0+
\dot{\alpha}_0 \dot{\Phi}_0 +\left( 2 \ddot{\alpha}_0
-\frac{\dot{\alpha}_0^2 \ddot{\alpha}_0}{\alpha_1^2} +3 \dot{\alpha}_0^2
\right) \Phi_0 \nonumber\\ 
&+& \frac{1}{3}e^{-2 \alpha_0}\nabla^2 \Phi_0-\frac{1}{3}
\left(1-\frac{\ddot{\alpha}_0}{\alpha_1^2} \right)
e^{-2 \alpha_0}\nabla^2 \Psi_0 
-\frac{\kappa^2 \alpha_1}{2} \delta p^{(4)}
+ \frac{\kappa^2}{2 l}
\delta p_{\chi},
\nonumber\\
-\frac{\kappa^2 \alpha_1}{2} (\rho+p) e^{\alpha_0}v &=&
\dot{\Psi}_0 +\dot{\alpha}_0 \Phi_0 
+\frac{1}{3} \alpha_1 e^{-3 \alpha_0} \int dt'
 e^{\alpha_0} \alpha_1^{-1} \left( \nabla^2 \Phi_0-\left(1-
\frac{\ddot{\alpha}_0}{\alpha_1^2} \right) \nabla^2 \Psi_0
\right) \nonumber\\
&-& \frac{\kappa^2 \alpha_1}{2} (\rho+p) e^{\alpha_0}v^{(4)} +
\frac{\kappa^2}{2 l} (\rho+p)e^{\alpha_0} v_{\chi},
\label{2-9}
\end{eqnarray}
where $\delta \rho^{(4)}$, $\delta p^{(4)}$ and $v^{(4)}$ 
satisfy
\begin{eqnarray}
\dot{\delta \rho}^{(4)}-3 \dot{\alpha}_0
(\delta \rho^{(4)}+\delta p^{(4)})&=&-\frac{2}{3 \kappa^2}e^{-5 \alpha_0}
\int dt' e^{\alpha_0} \alpha_1^{-1} \left(
\nabla^4 \Phi_0-\left(1-\frac{\ddot{\alpha}_0}{\alpha_1^2}\right)
\nabla^4 \Psi_0\right), \nonumber\\
\delta p^{(4)} &=&
\frac{1}{3}\left(1+\frac{\ddot{\alpha}_0}{\alpha_1^2} \right)
\delta \rho^{(4)}, \nonumber\\
\left( (\rho+p) e^{\alpha_0} v^{(4)} \right)^{\cdot}
&=&  -3 \dot{\alpha}_0 (\rho+p) e^{\alpha_0} v^{(4)} + \delta p^{(4)},
\label{2-10}
\end{eqnarray}
and $\delta \rho_{\chi}$, $\delta p_{\chi}$ and
$v_{\chi}$ are given by
\begin{eqnarray}
\delta \rho_{\chi} &=& e^{-4 \alpha_0} \chi(t,x^i),\nonumber\\
\delta p_{\chi} &=& \frac{1}{3} \left(1+
\frac{\ddot{\alpha}_0}{\alpha_1^2} \right)e^{-4 \alpha_0}
\chi(t,x^i),
\nonumber\\
(\rho+p)e^{\alpha_0} 
\nabla^2 v_{\chi} &=& e^{-2 \alpha_0} \dot{\chi}(t,x^i),
\label{2-11}
\end{eqnarray}
where $\chi$ satisfies (\ref{2-8}). 

(\ref{2-9}), (\ref{2-10}) and (\ref{2-11}) are the effective 4D Einstein 
equations in the brane world. 
The crucial difference from the background case is that we
cannot have a complete set of the equations. We do not have the
equation which corresponds to the $(i \neq j)$ component of the 
conventional 4D Einstein equations $\Phi_0=\Psi_0$. To clarify the 
deviation from the conventional 4D theory, we consider the 
perturbations at low energies with $\rho/\sigma \ll 1$, then
take
\begin{equation}
\alpha_1^2 \gg \dot{\alpha}_0^2, \:\: \ddot{\alpha}_0.
\end{equation}
If we take $\Phi_0=\Psi_0$, we have a complete set of the 
equations with (\ref{2-3}), (\ref{2-4}) and (\ref{2-5})
which are identical with those obtained in the conventional
4D theory. The interesting point is that even if we take 
$\Psi_0=\Phi_0$, 
the corrections to the matter perturbations can exist. 
Taking $\Psi_0=\Phi_0$, we find $\delta \rho^{(4)}$, 
$\delta \rho^{(4)}$ and $v^{(4)}$ and the higher order solutions
satisfy the homogeneous equations (\ref{2-6})
which do not include $\Psi_0$ and $\Phi_0$. Thus
they can be absorbed into $\delta \rho_{\chi}$,
$\delta p_{\chi}$ and $v_{\chi}$. Then 
(\ref{2-9}) becomes the same as
the conventional 4D Einstein equation except for $\delta \rho_{\chi}$,
$\delta p_{\chi}$ and $v_{\chi}$. Thus, even though we have 
a complete set of equations for metric perturbations and
matter perturbations which are identical with those obtained in 
the conventional 4D theory, the corrections to the 4D effective 
Einstein equation can exist. We have already learned the similar
situations in the background spacetime where the non-zero constant 
of the integration 
$C_0$ gives the correction to Friedmann equation. 
For the perturbations, $\chi$ plays the same role as $C_0$.  
At low energies, the equation for $\chi$ (\ref{2-8}) becomes
\begin{equation}
\quad \chi'' - \frac{1}{3} \nabla^2 \chi=0,
\label{2-12}
\end{equation}
where $'$ denotes the derivative with respect to the conformal time $\eta$.
At large scales and at low energies, $\delta \rho_{\chi}$ is given by
\begin{equation}
\delta \rho_{\chi}=C e^{-4 \alpha_0},
\end{equation}
where $\chi=C=const.$ 
Thus $\delta \rho_{\chi}$ can be regarded as the perturbations of 
the energy density of the 
dark radiation. At small scales they behave as sound waves with sound
velocity $1/\sqrt{3}$. 

In general, the bulk gravitational field makes $\Psi_0 \neq \Phi_0$
then the corrections arise which come from the special solutions written by
$\Phi_0$ and $\Psi_0$.  
Hence we find two types of the corrections to the matter perturbations.
One type of the corrections is given by the gradient of the metric 
perturbations. At large scales, these corrections are suppressed. 
These corrections are induced by the bulk perturbations.
The inhomogeneous matter perturbations on the brane inevitably produce the 
perturbations in the geometry of the bulk. The perturbations in the bulk 
affect the brane in turn. Then the matter perturbations receive the 
corrections from the bulk perturbations. 
Another type of the corrections, $\chi$, 
is independent of the metric perturbations. At 
large scales they behave as the dark radiation.  The 
corrections from $\chi$ are also induced by the bulk gravitational
field, as the dark radiation is induced by the Schwartzshild Black Hole 
in the bulk in the background spacetime. 

Now we face the limitation of the method using the equations solely on 
the brane. We cannot determine the corrections from the bulk perturbations, 
that is the relation between $\Phi_0$ and $\Psi_0$. The existence of the 
correction given by $\chi$ cannot also be determined, 
as the constant of the 
integration $C_0$ in the background 
cannot be determined in this approach. 
They should be determined in terms of the bulk gravitational field. 
So far we treated only the equations which do not involve the second 
derivative with
respect to $y$. As we showed for the background case, 
the evolution equation for the perturbations in the bulk should be solved 
in order to obtain the behaviour of the corrections to the 
matter perturbations and the relation between
$\Phi_0$ and $\Psi_0$. 


\section{Effective Einstein equations from 
bulk gravitational field}
\setcounter{equation}0
\hspace{0.5cm}
In this section, we solve the perturbations in the bulk 
and obtain the behaviour of the corrections 
to the matter perturbations and the relation between $\Psi_0$ and $\Phi_0$.
The formalism to solve the perturbations in the bulk 
has been developed in \cite{KJ}. In this section,
we only show the results of the calculations. 
The detailed calculations are given in the Appendix B.
In the bulk, the perturbations satisfy the wave equation
\begin{equation}
h''+ 3 \alpha' h'-\ddot{h} -3 \dot{\alpha} \dot{h}
+e^{-2 (\alpha-\beta)} \nabla^2 h=0,
\end{equation}
where $h$ is the scalar perturbations in the bulk and
we used the transverse-traceless gauge. It is in 
general difficult to solve the equation. The essence of our
method is to use the coordinate transformation from the 
Poincare coordinate to the Gaussian normal coordinate. 
The metric ((\ref{1-1}), (\ref{1-9}))
is obtained by the coordinate
transformation from the Poincare coordinate of the 5D AdS spacetime
\begin{equation}
ds^2= \left( \frac{l}{z} \right)^2
(dz^2- d\tau^2 +\delta_{ij}dx^{i}dx^{j}).
\end{equation}
In this coordinate, the perturbations can be easily
solved. Then the perturbations in the metric 
(\ref{1-1}) can be
obtained by performing the coordinate transformation;
\begin{eqnarray}
z&=&z(y,t)=l(f(u)-g(v))=l e^{-\alpha(y,t)}, \nonumber\\
\tau&=&\tau(y,t)=l(f(u)+g(v)).
\end{eqnarray}
The solution of the perturbations $h$ can be written as
\begin{equation}
h= e^{-2\alpha(y,t)} \int \frac{d^3 \k}{(2 \pi)^3}
\int d m \:\: 
E(m,\k) Z_2(ml e^{-\alpha(y,t)}) 
e^{-i \omega \tau(t,y)} e^{i \k \x}, 
\label{3-4}
\end{equation}
where $Z_2$ is defined as the combination of the Hankel function
of the first kind and the second kind;
\begin{equation}
 Z_2(m z)=H^{(1)}_2(mz) + a(m) H^{(2)}_2(m z),
\end{equation}
and $\omega^2=m^2+\k^2$. So far 
$E(m,\k)$ and $a(m)$ are arbitrary coefficients. 
We should impose the junction conditions on the perturbations 
(\ref{3-4}) on the brane.
As we showed for the background case, the junction conditions are
nothing but the effective 4D Einstein equations. 
In the previous paper \cite{KJ}, we gave the matter perturbations 
in terms of $E(m,\k)$ using the junction conditions. We also gave the
metric perturbations in terms of $E(m,\k)$. Then we have the equations
which correspond to (\ref{1-10}) and (\ref{1-11}) in the background. 
The effective Einstein equations can be obtained by combining 
these equations as is done in (\ref{1-12}) in the background spacetime.
The details can be found in Appendix B-2 and the results are 
given by 
\begin{eqnarray}
-\frac{\kappa^2 \alpha_1}{2} \delta \rho (\k)
&=& -3 (\dot{\alpha}_0 \dot{\Psi}_0 + \dot{\alpha}_0^2 \Phi_0)
-e^{-2 \alpha_0} \k^2 \Psi_0 \nonumber\\
&+&  
\frac{1}{3}e^{-4 \alpha_0} \int dm E(m,\k) \k^4 l^2
Z_0(m l e^{-\alpha_0}) e^{-i \omega T(t)}, \nonumber\\
-\frac{\kappa^2 \alpha_1}{2} \delta p(\k)
&=& \ddot{\Psi}_0+\left(3 \dot{\alpha}_0-\frac{\dot{\alpha_0} 
\ddot{\alpha}_0}{\alpha_1^2}\right) \dot{\Psi}_0+
\dot{\alpha}_0 \dot{\Phi}_0 +\left( 2 \ddot{\alpha}_0
-\frac{\dot{\alpha}_0^2 \ddot{\alpha}_0}{\alpha_1^2} +3 \dot{\alpha}_0^2
\right) \Phi_0 \nonumber\\
&-& \frac{1}{3}e^{-2 \alpha_0} \k^2 \Phi_0+\frac{1}{3}
\left(1-\frac{\ddot{\alpha}_0}{\alpha_1^2} \right)
e^{-2 \alpha_0}\k^2 \Psi_0 \nonumber\\
&+& \frac{1}{9}
\left(1+\frac{\ddot{\alpha}_0}{\alpha_1^2} \right)
e^{-4 \alpha_0} \int dm E(m,\k) \k^4 l^2
Z_0(m l e^{-\alpha_0}) e^{-i \omega T(t)}, \nonumber\\
-\frac{\kappa^2 \alpha_1}{2} (\rho+p) e^{\alpha_0}v(\k) 
&=&
\dot{\Psi}_0 +\dot{\alpha}_0 \Phi_0 \nonumber\\
&+& \frac{1}{3} e^{-3\alpha_0} \int dm E(m,\k) \nonumber \\
&& \times 
\left( 
\alpha_1 i \omega \k^2 l^3 Z_0 
(m l e^{-\alpha_0}) -\dot{\alpha}_0 
m \k^2 l^3  Z_1(m l e^{-\alpha_0})
\right) e^{-i \omega T(t)},
\label{3-6}
\end{eqnarray}
where we consider the Fourier components of the perturbations
with respect to $x^i$ and denote $\tau(0,t)=T(t)$. 
We can also obtain the metric perturbations in terms of 
$E(m,\k)$ as \\
\begin{eqnarray}
\Psi_0(\k) &=& \int dm E(m,\k)\left(
m l e^{-\alpha_0} Z_1 (m l e^{-\alpha_0})
+ \frac{1}{3} (\k l e^{-\alpha_0})^2 Z_0(m l 
e^{-\alpha_0})\right) e^{-i \omega T(t)}, \nonumber\\
\Phi_0(\k)&=& 
\int dm E(m,\k) 
\left(
m l e^{-\alpha_0} Z_1 (m l e^{-\alpha_0})
-\frac{1}{3}(\k^2+3 m^2)l^2 e^{-2 \alpha_0}
Z_0(m l e^{-\alpha_0}) \right) e^{-i \omega T(t)}
\nonumber\\
&+&(\dot{\alpha}_0 l)^2 \int dm E(m,\k)
\left(mle^{-\alpha_0} Z_1(m l e^{-\alpha_0}) 
- (\k^2+2 m^2)l^2  e^{-2 \alpha_0}
Z_0(m l e^{-\alpha_0}) \right)e^{-i \omega T(t)} \nonumber\\
&-& 2 \alpha_1 \dot{\alpha}_0 l^2 \int dm E(m,\k) 
(i \omega m l^2 e^{-2 \alpha_0}) Z_1(m l e^{-\alpha_0}) 
e^{-i \omega T(t)}.
\label{3-7}
\end{eqnarray} 

 Eqs.(\ref{3-6}) should be compared with (\ref{2-9}), (\ref{2-10}) and 
(\ref{2-11}). First let us identify the corrections 
$\delta \rho_{\chi}$, $\delta p_{\chi}$ and 
$v_{\chi}$ (\ref{2-11}). 
Using the freedom of the choice $E(m,\k)$
and $a(m)$, we can make the 
perturbations which do not contribute to the metric
perturbations but do contribute to the matter perturbations.
In the low energy universe,
we can construct these perturbations explicitly. 
At low energies $\dot{\alpha}_0 l \ll 1$, if we choose 
$E(m,\k)$ to have a peak at 
\begin{equation}
2 \k^2+3 m^2=0,
\end{equation}
then the metric perturbations can be written as
\begin{equation}
\Psi_0=\Phi_0= \frac{1}{2}E^{(\chi)}(\k) (m_k l e^{- \alpha_0})^2
Z_2(m_k l e^{-\alpha_0}) e^{\frac{1}{\sqrt{3}}i \k \eta},
\end{equation}
where we used $T=-\eta$ at low energies where $\eta$ is
the conformal time (see Appendix B (\ref{B-7})) and 
$Z_2(z)=(2/z)Z_1(z)-Z_0(z)$. We denoted
$m_k =\sqrt{2/3} \k i$ and 
\begin{equation}
E^{(\chi)}(\k)=E(m_k,\k).
\end{equation}
At low energies, we can neglect the time dependence in
$m_k l e^{-\alpha_0}$ since
\begin{equation}
\frac{\frac{d}{dt} (m_k l e^{-\alpha_0} )}{\frac{d}{dt}
(e^{i \omega \eta})}
\sim \dot{\alpha_0} l \ll 1,
\end{equation}
where we used $d \eta/d t= e^{-\alpha_0}$. 
In order to ensure that these perturbations do not contribute
to the metric perturbations $\Phi_0=\Psi_0=0$,  
we take $a(m)=a^{(\chi)}(m)$ where
\begin{equation}
a^{(\chi)}(m)=-\frac{H^{(1)}_2(m_k l e^{-\alpha_0})}
{H^{(2)}_2 (m_k l e^{-\alpha_0})}.
\label{3-12}
\end{equation}
The important point is that these perturbations do 
contribute to the matter perturbations. 
The density perturbations $\delta \rho$ induced by these perturbations
are given by 
\begin{equation}
\frac{\kappa^2}{2 l}
\delta \rho
=  \frac{1}{3} e^{-4 \alpha_0} \k^4 l^2 
E^{(\chi)}(\k) Z^{(\chi)}_0 (m_k l e^{-\alpha_0}) 
e^{\frac{1}{\sqrt{3}} i \k \eta},
\label{3-13}
\end{equation}
where $Z^{(\chi)}_0=H^{(1)}_0+a^{(\chi)}(m) H^{(2)}_0$.
Because these perturbations do not contribute to the 
metric perturbations, they should be identified 
as $\delta \rho_{\chi}$.
Indeed from (\ref{2-7}) and (\ref{2-12}), $\delta \rho_{\chi}$
is given by 
\begin{equation}
\delta \rho_{\chi}= e^{-4 \alpha_0} \chi ,
\quad \chi'' +\frac{1}{3} \k^2 \chi=0.
\label{3-14}
\end{equation}
If we neglect the time dependence in $m_k l e^{-\alpha_0}$,
(\ref{3-13}) satisfies (\ref{3-14}).
Thus we find the existence of $\delta \rho_{\chi}$ depends on 
the behaviour of the bulk perturbations. At low energies, they should
have imaginary mass $m_k=\sqrt{2/3} \k i$ and diverge at the horizon of 
the AdS spacetime ($z= l e^{-\alpha(y,t)} \to \infty$) 
because $Z^{(\chi)}_2(m_k z)$ contains $H^{(2)}_2(m_k z)$
which is proportional to $\exp(\sqrt{2/3} \k z)$ for $z \to \infty$. 
Thus if we restrict our attention to the bulk perturbations with 
real mass or with regular behaviour in the bulk, the corrections 
from $\chi$ do not exist on the brane. 

Therefore the existence of the corrections from $\chi$ depends on the
boundary condition of the perturbations (\ref{3-4}).  
The general solutions for perturbations in the bulk can be written as
\begin{equation}
h(\k)=e^{-2 \alpha(y,t)} \int dm
\left(E^{(1)}(m,\k) Z^{(1)}(m l e^{-\alpha(y,t)}) +
E^{(2)}(m,\k) Z^{(2)}(m l e^{-\alpha(y,t)}) \right)
e^{-i \omega \tau(y,t)},
\label{3-15}
\end{equation}
where $Z^{(1)}$ and $Z^{(2)}$ are two
independent combinations 
of the Hankel function of the first kind and the second kind. 
$E^{(1)}(m,\k)$ and $E^{(2)}(m,\k)$ are the arbitrary
coefficients that should be determined by the boundary
conditions. One of the choices is the boundary condition that allows the
existence of the corrections $\delta \rho_{\chi}$.
We choose $E^{(2)}(m,\k)$ and $Z^{(2)}$ so that the perturbations 
contribute to the matter perturbations and do not contribute to the
metric perturbations. For example, at low energies, we can take
\begin{eqnarray}
Z^{(1)}(m l e^{-\alpha})&=& H^{(1)}(m l e^{-\alpha}), \nonumber\\
Z^{(2)}(m l e^{-\alpha}) &=& Z^{(\chi)}(m_k l e^{-\alpha}),\nonumber\\
E^{(2)}(m,\k)&=& E^{(\chi)}(\k).
\label{3-16-0}
\end{eqnarray}
Then the metric perturbations and the density perturbation induced by these
perturbations are given by
\begin{eqnarray}
\frac{\kappa^2}{2 l}
\delta \rho
&=& -3 (\dot{\alpha}_0 \dot{\Psi}_0 + \dot{\alpha}_0^2 \Phi_0)
-e^{-2 \alpha_0} \k^2 \Psi_0 \nonumber\\
&+&  \frac{1}{3} e^{-4 \alpha_0}
\int dm E^{(1)}(m,\k) \k^4 l^2
H_0^{(1)}(mle^{-\alpha_0})
e^{i \omega \eta}+ \frac{\kappa^2}{2 l} \delta \rho_{\chi}, \nonumber\\
\Psi_0(\k) &=& \int dm E^{(1)} (m,\k)\left(
m l e^{-\alpha_0} H^{(1)}_1 (m l e^{-\alpha_0})
+ \frac{1}{3} (\k l e^{-\alpha_0})^2 H^{(1)}_0(m l 
e^{-\alpha_0}) \right) e^{i \omega \eta},
\label{3-19}
\end{eqnarray}
where
\begin{eqnarray}
\delta \rho_{\chi} &=& e^{-4 \alpha_0} \chi, \nonumber\\
\chi &=& \frac{2 l}{3 \kappa^2}
(\k^4 l^2 E^{(\chi)}(\k))
Z_0^{(\chi)}(m_k l e^{-\alpha_0}) e^{\frac{1}{\sqrt{3}}i
\k \eta}.
\end{eqnarray}
Another choice is the boundary condition 
that the perturbations are out-going 
at the horizon of the AdS spacetime \cite{KJ},\cite{Haw},\cite{GKR}. 
Then we should take
\begin{eqnarray}
Z^{(1)}(m l e^{-\alpha})&=& H^{(1)}(m l e^{-\alpha}),\nonumber\\
E^{(2)}(m,\k)&=& 0.
\label{3-16}
\end{eqnarray}
Note that for imaginary $m=i m_I, \:\: m_I >0$, this condition
implies that the perturbations do not diverge at the horizon
of the AdS spacetime because $H^{(1)}(i m_I z) \propto
\exp(-m_I z)$ at $z \to \infty$. 
Hence if we take the boundary 
condition that the perturbations are out-going, the
corrections given by $\chi$ are not allowed; 
\begin{equation}
\delta \rho_{\chi}=\delta p_{\chi}=v_{\chi}=0.
\end{equation}
The matter perturbations and the metric perturbations 
are given by (\ref{3-6}) and (\ref{3-7}) with 
$E(m,\k)=E^{(1)}(m,\k)$ and $Z(m l e^{-\alpha_0})=
H^{(1)}(m l e^{-\alpha_0})$. 

It seems difficult to determine what kind of the perturbations
are allowed in the bulk. We will discuss the effects of the 
corrections from $\chi$ separately according to the 
choice of the boundary condition in section 5. For a while
we take the boundary condition (\ref{3-16})
and take $\delta \rho_{\chi}
=\delta p_{\chi}=v_{\chi}=0$.
The terms written in terms of $E^{(1)}(m,\k)$ in the matter perturbations 
(\ref{3-6}) correspond to the 
corrections written by the gradient of the metric perturbations 
in (\ref{2-9}) and (\ref{2-10}). 
In fact, if we take $\k \to 0$, the terms written in $E^{(1)}(m,\k)$
in the matter perturbations vanish. Now from (\ref{3-6}) and 
(\ref{3-7}), we manifestly observe that the bulk perturbations alter
the relation $\Psi_0=\Phi_0$ and induce the corrections to the matter
perturbations. In (\ref{3-15}), 
$E^{(1)}(m,\k)$ is still the arbitrary coefficient. 
$E^{(1)}(m,\k)$ is determined once we impose the
equation of state of the matter perturbations such as 
$\delta p= c_s^2 \delta \rho$ where $c_s^2$ is the sound velocity. 
It is rather difficult to solve the equations for $E^{(1)}(m,\k)$. 
In the following section, we try to obtain the evolution of the 
perturbations without solving $E^{(1)}(m,\k)$. The price to pay is 
that we should take an assumption about the contribution from the
massive modes as in \cite{KJ}. 

From the effective Einstein equations (\ref{3-6}) and
(\ref{3-7}), we find there are 
two situations in which the deviation from the conventional 
4D theory becomes large. One is given by
\begin{equation}
\k l e^{-\alpha_0} \gg1, 
\end{equation}
that means the physical scale of the perturbations is smaller
than the curvature scale $l$. It is reasonable since $l$ is the effective
scale of the compactification, thus at the scales smaller than $l$,
the gravity behaves as the 5D one. Another is given by
\begin{equation}
\dot{\alpha_0} l \gg 1,
\end{equation}
that means the energy density of the matter exceeds the tension
of the brane. In the Friedmann equation (\ref{1-12}), 
the term proportional to 
$\rho^2$ becomes
dominant and the evolution of the universe changes significantly.


\section{Modifications of the evolution}
\setcounter{equation}0
In this section, we take the boundary condition that the
perturbations are out-going at the horizon of the AdS spacetime
(\ref{3-16}).
Then
\begin{equation}
\delta \rho_{\chi}=\delta p_{\chi}=v_{\chi}=0.
\end{equation}
In the following sections, 
we assume the matter perturbations are adiabatic.

\subsection{Evolution at super-horizon scales}
\hspace{0.5cm}
Let us consider the long-wave length perturbations. We take
\begin{equation}
\k l e^{-\alpha_0} \to 0, 
\end{equation}
then the corrections to the matter perturbations in (\ref{3-6})
written by $E^{(1)}(m,\k)$ vanish.
The evolution equation for the metric perturbations is given
from (\ref{3-6}) by imposing $\delta p-c_s^2 \delta \rho=0$. 
The equation can be simplified using the Bardeen parameter
\begin{equation}
\zeta=\Psi_0-\frac{\dot{\alpha}_0^2}{\ddot{\alpha}_0} \left(
\frac{1}{\dot{\alpha}_0} \dot{\Psi}_0+\Phi_0 \right).
\label{4-3}
\end{equation}
At the super-horizon scales $\k \dot{\alpha}_0^{-1}e^{-\alpha_0} \ll 1$, 
$\delta p - c_s^2 \delta \rho=0$ can be written as
\begin{equation}
\dot{\zeta}=0,
\end{equation}
where we used
\begin{equation}
\dot{w}=-3 \dot{\alpha_0}(1+w)(1-c_s^2).
\label{4-4}
\end{equation}
Then the Bardeen parameter is conserved even at high energies
\begin{equation}
\zeta=\zeta_{\ast}=const.
\label{4-5}
\end{equation}
We should note that the constancy of the Bardeen parameter does not
mean that the behaviour of the perturbations in the brane world is 
the same as the one obtained in 4D theory.
The Bardeen parameter is written by $\Phi_0$ and $\Psi_0$. In the 4D
theory we have the equation $\Phi_0=\Psi_0$. 
However in the brane world 
it is modified by the perturbations in the bulk.
The equation that gives the relation between $\Psi_0$ and $\Phi_0$ is needed. 
From (\ref{3-7}), the metric perturbations are given by
\begin{eqnarray}
\Psi_0 &=&\int dm E^{(1)}(m)m l e^{-\alpha_0} 
H^{(1)}_1(m l e^{-\alpha_0}) e^{-i m T},\nonumber\\
\Phi_0 &=& (1+(\dot{\alpha}_0 l)^2)\Psi_0
-(1+2 (\dot{\alpha}_0 l)^2)
\int dm E^{(1)}(m) 
(m l e^{-\alpha_0})^2 H^{(1)}_0(ml e^{-\alpha_0})e^{-i m T} 
\nonumber\\
&-&  2 i \alpha_1 \dot{\alpha}_0 l^2 
\int dm E^{(1)}(m)(m l e^{-\alpha_0})^2  
H^{(1)}_1(m l e^{-\alpha_0})e^{- i m T}.
\end{eqnarray}
As mentioned in the previous section, we should make 
some assumption about the contribution from massive modes.
We will assume that the modes with $m le^{-\alpha_0}
>1$ do not contribute to the perturbations in the bulk thus take
\begin{equation}
m l e^{-\alpha_0} \to 0.
\end{equation}
Then using the asymptotic form of the Hankel function
$H^{(1)}_1(z) \propto 1/z$ and $H^{(1)}_0(z) \propto const.$, 
we obtain
\begin{eqnarray}
\Phi_0 &=& (1+ (\dot{\alpha}_0 l)^2) \Psi_0.
\end{eqnarray}

At high energies, we have
\begin{equation}
\Phi_0 = (\dot{\alpha}_0 l)^2 \Psi_0,
\end{equation}
then $\Phi_0 \gg \Psi_0$.
From the conservation of the Bardeen 
parameter (\ref{4-3}) and (\ref{4-5}), we get
\begin{eqnarray}
\Phi_0 &=& 3(1+w) \zeta_{\ast} ,\quad \Psi_0=(\dot{\alpha}_0 l)^{-2} 
\Phi_0,
\nonumber\\
\frac{\delta \rho}{\rho} &=& - \Phi_0,
\end{eqnarray}
for $w=const.$
Note that the curvature perturbation increases as $ \Psi_0 
\propto \rho^{-2}$ at high energies. 

At low energies $\dot{\alpha}_0 l \ll 1$, we have
\begin{equation}
\Psi_0=\Phi_0.
\end{equation}
Then the metric perturbations are obtained as
\begin{eqnarray}
\Phi_0 &=& \Psi_0=\frac{3(1+w)}{5+3 w} \zeta_{\ast}, \nonumber\\
\frac{\delta \rho}{\rho} &=& -2 \Phi_0,
\label{4-10}
\end{eqnarray} 
for $w=const.$ 

The CMB anisotropies at large scales can be obtained 
using the above solutions. At the
decoupling of the photon and baryon, the energy of the
universe is lower than the tension of the brane $\dot{\alpha}_0 l \ll 1$.  
The temperature anisotropies caused by the ordinary Sachs-Wolfe
effect are given by
\begin{equation}
\frac{\Delta T}{T}=\frac{1}{4}\frac{\delta \rho_r}{\rho_r}
+\Phi_0=\frac{1}{3}\frac{\delta \rho}{\rho}+\Phi_0,
\end{equation}
where $\rho_r$
is the density of the radiation and $\delta \rho_r$ is its
perturbation. From (\ref{3-6}) and (\ref{4-3}), 
we can show the Bardeen parameter is given by
\begin{equation}
\zeta=\Psi_0 -\frac{1}{3}\frac{\delta \rho}{\rho}.
\end{equation}
Then the temperature anisotropies can be evaluated as
\begin{equation}
\frac{\Delta T}{T}=-\zeta+\Psi_0+\Phi_0.
\label{4-15}
\end{equation}
If we neglect the effect of the massive graviton with
$m l e^{-\alpha_0}>1 $, we can 
evaluate the temperature anisotropies as
\begin{equation}
\frac{\Delta T}{T}=\frac{1}{5} \zeta_{\ast},
\label{4-16}
\end{equation}
where we used the solution (\ref{4-10}) with $w=0$.

The massive graviton will modify the relation
$\Phi_0$ and $\Psi_0$ so the temperature anisotropies.
At low energies, the metric perturbations are given by
\begin{eqnarray}
\Psi_0 &=&\int dm E^{(1)}(m)m l e^{-\alpha_0} 
H^{(1)}_1(m l e^{-\alpha_0}) e^{i m \eta},\nonumber\\
\Phi_0 &=& \Psi_0
-\int dm E^{(1)}(m) 
(m l e^{-\alpha_0})^2 H^{(1)}_0(ml e^{-\alpha_0})e^{i m \eta}. 
\end{eqnarray}
Then at the lowest order corrections in $ml e^{-\alpha_0}$
we have
\begin{equation}
\Phi_0= \Psi_0 - \int dm \Psi_0(m) (ml e^{-\alpha_0})^2
G_{KK}(m l e^{-\alpha_0}) e^{i m \eta}, 
\label{4-18}
\end{equation}
where $\Psi_0(m)$ denotes the
Fourier transformation of $\Psi_0(\eta)$ with respect to $\eta$
and 
\begin{eqnarray}
G_{KK}(mle^{-\alpha_0}) &=&  \lim_{m l e^{-\alpha_0} \to 0} 
\left(\frac{H^{(1)}_0(m l e^{-\alpha_0} )}
{m l e^{-\alpha_0} H^{(1)}_1(m l e^{-\alpha_0})}\right)
\nonumber\\
&=& \ln(2 e^{\alpha_0})-\gamma+\frac{\pi}{2}i-\ln(ml),
\label{4-19}
\end{eqnarray}
where $\gamma$ is the Euler number. The important 
point is that $G_{KK}$ contains non analytic  
term proportional to $\ln m$. Then (\ref{4-18}) becomes non-local 
when we make Fourier transformation
to the real spacetime. The reason can be understood as 
follows. The massive modes with $m \neq 0$ can propagate into
the bulk. These modes affect the
metric perturbations non-locally if they are observed
on the brane. Then the non-locality of the evolution equation
is the essential feature of the brane world \cite{Muko}. 

The contributions from the massive modes are determined by
$E^{(1)}(m)$ which is determined by the primordial 
fluctuations and later evolutions. It is difficult to 
know $E^{(1)}(m)$, but it should be noted that in (\ref{4-18}), $m$ 
appears in the form $m l e^{-\alpha_0}$. Thus as the
energy of the universe become lower $e^{-\alpha_0} \to 0$, 
the mass of the massive modes that can modify the relation 
$\Phi_0=\Psi_0$ becomes larger. Then for late times, 
we can safely use the standard result (\ref{4-16}).
The constant $\zeta_{\ast}$ should be determined
by the primordial fluctuations. We discuss the 
generation of the primordial fluctuations in the
Appendix C. 

Here is the point we should emphasize. At high energies,
the Hubble scale itself is smaller than the curvature
scale of the AdS spacetime. Then we should be careful in
using the result $\k l e^{-\alpha_0}  \to 0$ even at
the super-horizon scales in the high energy universe.

\subsection{Evolution at sub-horizon scales}
\hspace{0.5cm}
In this section we investigate the corrections which arise for
\begin{equation}
\k l e^{-\alpha_0} \neq 0.
\end{equation}
We will assume the universe is in the low energy era 
$\dot{\alpha}_0 l \ll 1$ and take
\begin{equation}
m l e^{-\alpha_0} \to 0.
\end{equation}
At the sub-horizon scales $\k \dot{\alpha}_0^{-1} 
e^{-\alpha_0} \gg 1$, 
the density perturbation (\ref{3-6}) is given by 
\begin{eqnarray}
\frac{\kappa^2}{2 l} e^{2 \alpha_0} \delta \rho
&=& -\k^2 \Psi_0 + \frac{1}{3} e^{-2 \alpha_0} 
\int dm E^{(1)}(m,\k) \k^4 l^2 H^{(1)}_0(m l e^{-\alpha_0}) 
e^{i \omega \eta} \nonumber\\
&=&- \int dm E^{(1)}(m,\k) \k^2 
m l e^{-\alpha_0} H^{(1)}_1(m l e^{-\alpha}_0)
e^{i \omega \eta}.
\end{eqnarray}
On the other hand the metric perturbations (\ref{3-7}) are given by
\begin{eqnarray}
\k^2 \Phi_0 &=& 
\int dm E^{(1)}(m,\k) 
\k^2 \left(
m l e^{-\alpha_0} H^{(1)}_1 (m l e^{-\alpha_0})
-\frac{1}{3}(\k^2+3 m^2)l^2 e^{-2 \alpha_0}
H^{(1)}_0(m l e^{-\alpha_0}) \right) e^{i \omega \eta}. \nonumber\\
\label{4-23}
\end{eqnarray}
For $m l e^{-\alpha_0} \ll 1$ we can rewrite (\ref{4-23}) 
into the effective Poisson equation;
\begin{equation}
\k^2 \Phi_0= -\frac{\kappa^2}{2 l} e^{2 \alpha_0}
\delta \rho+ \frac{\kappa^2 l}{6} 
e^{-2 \alpha_0} \int dm (\k^2+3 m^2)
G_{KK}(m l e^{-\alpha_0}) \left[e^{2 \alpha_0} \delta \rho
\right](m,\k),
\label{4-24}
\end{equation}
where $G_{KK}$ is given by (\ref{4-19})
and $[e^{2 \alpha_0} \delta \rho](m,\k)$ denotes the
Fourier transformation of $e^{2 \alpha_0} \delta \rho$ 
with respect to $\eta$ and $x^i$. 
For $\k l e^{-\alpha_0}
\to 0$, (\ref{4-24}) is the usual Poisson equation.
The evolution equation for $\delta \rho$
can be derived from the conservations law
of the matter perturbations (\ref{2-4}) and (\ref{2-5}).
For example, in the matter dominated era $w=0$, we get
\begin{eqnarray}
\lefteqn{\Delta'' + \alpha_0' \Delta'
-\frac{3}{2} \alpha_0'^2 \Delta}
\nonumber\\
&& + \frac{\kappa^2 l}{6}e^{-2 \alpha_0}
\int dm (\k^2 +3 m^2)G_{KK}(m l e^{-\alpha_0})
\left[e^{2 \alpha_0} \delta \rho \right](m,\k) e^{-i \omega \eta}=0,
\end{eqnarray}
where $'$ denotes the derivative with respect to $\eta$ 
and $\Delta=\delta \rho/ \rho$. 
The last term represents
the correction from the bulk perturbations. 
Note that for $\k l e^{-\alpha_0} \neq 0$, the non-local term
arises even if we take $ml e^{-\alpha_0} \to 0$.
This is because the graviton can easily propagate into the 
bulk at the scales smaller than $l$ ($\k l e^{-\alpha_0} >1$). 
Thus the bulk gravitational field affects the evolution of the
density perturbation non-locally. 

It is well known that in Minkowski spacetime, the Newton's low
is modified due to the 5D graviton [3,31-33]
This modification should be
derived from the effective Poisson equation. 
Let us consider the situation where $e^{\alpha_0}=1$.  
We assume that the source is static $\omega^2=m^2+\k^2=0$ and
derive the lowest order corrections in $(\k l)^2<1$.
Taking the non-analytic term in $m$, the metric perturbations
are written as
\begin{eqnarray}
\Psi_0(\k)&=& -\frac{\kappa^2}{2l}\left(\k^{-2}-
\frac{1}{3} l^2 \ln(m l) \right)\delta \rho(\k),\nonumber\\
\Phi_0(\k)&=& -\frac{\kappa^2}{2l}\left(\k^{-2}-
\frac{2}{3} l^2 \ln(m l) \right)\delta \rho(\k).
\end{eqnarray}
To compare the result with the one obtained in [31-33], we consider
the spherically symmetric source and derive the metric perturbations
far away from the source. We obtain the metric perturbations by 
the Fourier transformation as
\begin{eqnarray}
\Psi_0(r)= -\frac{G_4 M}{r} \left(1+ \frac{l^2}{3 r^2} \right),
\nonumber\\
\Phi_0(r)= -\frac{G_4 M}{r} \left(1+ \frac{2 l^2}{3 r^2} \right),
\end{eqnarray}
where $8 \pi G_4=\kappa^2/l$, $M=\int d \x^3 \delta \rho(x)$ 
and the source is located at $r=0$.
The result completely agrees with the one obtained in [31-33]. 


\section{Corrections from  
$\delta \rho_{\chi}$, $\delta p_{\chi}$ and $v_{\chi}$}
\setcounter{equation}0
In this section we choose the boundary condition 
so that the corrections $\delta \rho_{\chi}$, $\delta p_{\chi}$ 
and $v_{\chi}$ are induced on the brane. Then we investigate
the effects of the corrections on the evolution of the 
perturbations. 

\subsection{Evolution at super-horizon scales}
\hspace{0.5cm}
Let us consider the long wave-length perturbations 
with $\k l e^{-\alpha_0} \to 0$.  The corrections 
to the matter perturbations given by $E^{(1)}(m,\k)$ vanish. 
At the super-horizon scales $\k \dot{\alpha}_0^{-1} e^{-\alpha_0} 
\ll 1$, the 
density perturbation and the pressure perturbation
are obtained from (\ref{2-11}) and (\ref{3-6}) as 
\begin{eqnarray}
-\frac{\kappa^2 \alpha_1}{2} \delta \rho
&=& -3 (\dot{\alpha}_0 \dot{\Psi}_0 + \dot{\alpha}_0^2 \Phi_0)+
\frac{\kappa^2}{2 l} C e^{-4 \alpha_0}, \nonumber\\
-\frac{\kappa^2 \alpha_1}{2} \delta p
&=& \ddot{\Psi}_0+\left(3 \dot{\alpha}_0-\frac{\dot{\alpha_0} 
\ddot{\alpha}_0}{\alpha_1^2}\right) \dot{\Psi}_0+
\dot{\alpha}_0 \dot{\Phi}_0 +\left( 2 \ddot{\alpha}_0
-\frac{\dot{\alpha}_0^2 \ddot{\alpha}_0}{\alpha_1^2} +3 \dot{\alpha}_0^2
\right) \Phi_0 \nonumber\\ 
&& + \frac{\kappa^2}{2l}C e^{-4 \alpha_0}
\frac{1}{3} \left(1+
\frac{\ddot{\alpha}_0}{\alpha_1^2} \right),
\end{eqnarray}
where $\chi=C=const.$ from (\ref{2-8}). 
Then using the Bardeen parameter (\ref{4-3}), 
$\delta p - c_s^2 \delta \rho=0$ can be written as
\begin{equation}
\dot{\zeta}=  \frac{\kappa^2}{2 l}
C e^{-4 \alpha_0}\frac{\dot{\alpha}_0}{\ddot{\alpha}_0}
\left(\frac{1}{3} \left(1+ \frac{\ddot{\alpha}_0}{\alpha_1^2}
\right)-c_s^2 \right).
\end{equation}
We see the term proportional to $C$ breaks the constancy of the
Bardeen parameter. The results can be understood as follows.
The density perturbation $\delta \rho_{\chi}$ induce isocurvature
perturbations on the brane.
In 4D theory, it is well known that the isocurvature perturbations
break the constancy of the Bardeen parameter. In fact, if we 
consider the perturbations in radiation dominated era at low
energies
\begin{equation}
c_s^2=\frac{1}{3},\:\:\: \alpha_1 \gg \dot{\alpha}_0,
\end{equation}
then $\dot{\zeta}=0$. 
It is reasonable since $\delta \rho_{\chi}$ behaves as 
radiation ($\delta p_{\chi}=(1/3) \delta \rho_{\chi}$) at
low energies, so there are no isocurvature perturbations. 
The equation can be integrated using the background equations
(\ref{A-12}) and (\ref{4-4}).
We get
\begin{eqnarray}
\zeta&=&
\zeta_{\ast}-\frac{\kappa^2}{2 l}
\frac{C e^{-4 \alpha_0}}{3 \ddot{\alpha}_0}\nonumber\\
&=&
\zeta_{\ast}- \frac{1}{3(1+w)} \frac{1}{l \alpha_1}
\left(\frac{\rho_r}{\rho} \right) C_{\ast},
\end{eqnarray}
where we defined
\begin{equation}
C_{\ast}
=C \frac{e^{-4 \alpha_0}}{\rho_r}= const.
\label{5-5}
\end{equation}
Using the expression of the Bardeen parameter 
in terms of the metric perturbations (\ref{4-3}), we can obtain
the solutions of the metric perturbations.

At high energies, using
\begin{equation}
\Psi_0 =\frac{1}{(\dot{\alpha}_0 l)^2} \Phi_0 \ll \Phi_0,
\end{equation}
we get
\begin{equation}
\Phi_0=3(1+w) \zeta_{\ast}+\frac{1}{\dot{\alpha}_0 l} 
\left(\frac{\rho_r}{\rho} \right)
 C_{\ast},
\label{5-7}
\end{equation}
for $w=const.$
Note that the contribution of $C_{\ast}$ is 
suppressed by the factor $(\dot{\alpha}_0 l)^{-1}$. 

At low energies, using
\begin{equation}
\Psi_0=\Phi_0,
\end{equation}
we get
\begin{eqnarray}
\Phi_0 &=& \frac{3(1+w)}{5+3 w} \zeta_{\ast}
+\frac{1}{3(1+3 w)} \left(\frac{\rho_r}{\rho} \right)
C_{\ast},
\label{5-9}
\end{eqnarray}
for $w=const.$ 

The CMB anisotropies caused by the ordinary Sachs-Wolfe effect 
at low energy matter dominated era (\ref{4-15}) 
is given by
\begin{eqnarray}
\frac{\Delta T}{T} &=& - \zeta+2 \Phi_0 \nonumber\\
&=& \frac{1}{5} \zeta_{\ast}+\frac{1}{3} 
\left(\frac{\rho_r}{\rho} \right) C_{\ast}.
\end{eqnarray}
From the observations, $\Delta T/T \sim 10^{-5}$ and at
the decoupling $\rho_r/\rho \sim 0.1$. Then the constraint on 
$C_{\ast}$ is obtained as \cite{Large}
\begin{equation}
C_{\ast} < 10^{-4}. 
\end{equation}

\subsection{Evolution at sub-horizon scales}
\hspace{0.5cm}
Now consider the evolution of the perturbations at the
sub-horizon scales $\k \dot{\alpha}_0^{-1} 
e^{-\alpha_0} \gg 1$. For simplicity, we assume the universe is 
at low energies. We also assume the scale of the perturbations
is larger than $l$ ($\k l e^{-\alpha_0} \ll 1$). 
Then the corrections given by $E^{(1)}(m,\k)$ can be neglected.
In order to describe
the evolution of the density perturbation, it is convenient
to introduce the gauge invariant variable defined by
\begin{equation}
\rho \Delta=\delta \rho+3 \dot{\alpha}_0 (\rho+p)
e^{\alpha_0} v.
\end{equation}
From (\ref{3-6}) and (\ref{2-11}) 
the Poisson equation is given by
\begin{equation}
\k^2 \Phi_0=-\frac{3}{2}\alpha_0'^2 \Delta +e^{-2\alpha_0}
\frac{\kappa^2}{2 l}(\chi-3 \alpha_0' \k^{-2} \chi').
\label{5-13}
\end{equation}
The conservations of the matter (\ref{2-4}) and (\ref{2-5}) 
become
\begin{eqnarray}
\Delta'-3 w \alpha_0' \Delta &=& -(1+w) \k^2 v
+\frac{3 \kappa^2}{2 l}(1+w)e^{-2 \alpha_0} \k^{-2} \chi' ,\nonumber\\
v'+\alpha_0' v &=& \Phi_0+\frac{c_s^2}{1+w} \Delta,
\end{eqnarray}
where we used (\ref{4-4}).
Then the evolution equation for $\Delta$ can be obtained as
\begin{equation}
\Delta''-\left(3(2 w-c_s^2)-1 \right) \alpha_0' \Delta'
+3 \left(\frac{3}{2} w^2-4 w-\frac{1}{2}+3 c_s^2 \right)
\alpha_0'^2 \Delta+c_s^2 \k^2 \Delta=- 
\frac{\kappa^2}{l}(1+w)e^{-2 \alpha_0}
\chi.
\label{5-15}
\end{equation}
The initial condition of $\Delta$ can be set in the radiation
dominated era. In the radiation dominated era,
$w=c_s^2=1/3$ and $e^{\alpha_0}=\eta$, (\ref{5-15}) becomes 
\begin{equation}
\Delta''- \frac{2}{\eta^2} \Delta+\frac{1}{3} \k^2 \Delta
=-\frac{4 \kappa^2}{3 l} \frac{1}{\eta^2} \chi.
\end{equation}
Then we can easily find the solution as
\begin{equation}
\Delta=A U_G(\eta)+B U_D(\eta)+\frac{2 \kappa^2}{3 l} \chi,
\end{equation}
where
\begin{eqnarray}
U_G &=& -\cos \left(\k_s \eta \right)+
\left(\frac{1}{\k_s \eta} \right) \sin 
\left(\k_s \eta \right), \nonumber\\
U_D &=& -\sin \left(\k_s \eta \right)-
\left(\frac{1}{\k_s \eta} \right) \cos
\left(\k_s \eta \right),
\end{eqnarray}
and $A$ and $B$ are the constants of the integration and
$\k_s=\k/\sqrt{3}$. 
Note that for $\k \eta \to 0$, $U_G$ and $U_D$ behave as
$U_G =\k^2 \eta^2/9$ and $U_D \propto \eta^{-1}$. Then
$U_G$ matches to the growing mode solution at the super-horizon
scales. 
From (\ref{5-13}), The metric perturbation is given by
\begin{equation}
\Phi_0=- \frac{1}{\k^2  \eta^2} \left(
\frac{3}{2}(A U_G+B U_D) +\frac{\kappa^2}{2 l}(
\chi+\frac{3}{\eta} \k^{-2} \chi') \right).
\end{equation}
From (\ref{3-14}), the solution for $\chi$ is given by
\begin{equation}
\chi=C \cos \left(\k_s \eta \right) +D 
\sin \left(\k_S \eta \right),
\end{equation}
where $D$ is the arbitrary constant. 
Then $\Phi_0$ becomes
\begin{equation}
\Phi_0= -\frac{1}{\k^2 \eta^2} \left(
\left(\frac{3}{2} A - \frac{\kappa^2}{2 l}C \right)U_G +
\left(\frac{3}{2} B -\frac{\kappa^2}{2 l} D \right) U_D \right).
\end{equation}
We take only the growing mode solution, then $B=D=0$.
In the radiation dominated era at low energies, 
\begin{equation}
\alpha_0'^2 e^{-2 \alpha_0}= \frac{\kappa^2}{3 l} \rho_r,
\end{equation}
and $e^{\alpha_0}=\eta$. Then (\ref{5-5}) becomes 
\begin{equation}
C_{\ast}=\frac{\kappa^2}{3 l} C.
\end{equation}
At super-horizon scales, using $U_G = \k^2 \eta^2/9$, 
we get
\begin{equation}
\Phi_0=-\frac{A}{6}+\frac{1}{6} C_{\ast} .
\end{equation}
Comparing the solution with (\ref{5-9}), we should 
take $A = -4 \zeta_{\ast}$. Thus, we can set the
initial condition of $\Delta$ at radiation dominated era as
\begin{equation}
\Delta= -4 \zeta_{\ast} U_G+2 C_{\ast}
\cos (\k_s \eta).
\label{5-23}
\end{equation}
In the radiation dominated era, the perturbations are 
constant $\Delta \sim 2  C_{\ast}$
at super-horizon scales and then oscillate as cosine 
function once they enter the horizon. 
Thus at sub-horizon scales, the density perturbation
behaves as the usual adiabatic perturbations in 4D theory.
 However, as the matter becomes dominant, the isocurvature perturbations 
are generated. This is because while 
the frequency of the $U_G$ changes from $\k_s$,
$\chi$ always oscillates with frequency $\k_s$. Then there is 
a possibility that the amplitude and the phase of the oscillations of
$\Delta$ change from the 
adiabatic cosine mode. These deviations can be directly 
observed as the shifts of the location and height of the peak of 
the acoustic oscillation in CMB spectrum. We solve (\ref{5-15})
numerically with the initial condition given by (\ref{5-23}). 
In Fig.1 the density perturbation $\Delta(\k)$ at the time 
$\rho_r/\rho_m=0.1$ is shown with various wave numbers $\k$. 
where $\rho_m$ is the density of the matter.
For $C_{\ast}=0$, $\Delta(\k)$ 
is given by cosine function.
If we include the effect of $C_{\ast}$
the location and height of the peak of the oscillations 
change as expected.
Thus if we include the effect of the corrections 
$\delta \rho_{\chi}$, $\delta p_{\chi}$ and $v_{\chi}$, 
the effects from the bulk can be observed even in the low 
energy universe. 

\begin{figure}[ht]
  \epsfysize=90mm
\begin{center}
  \epsfbox{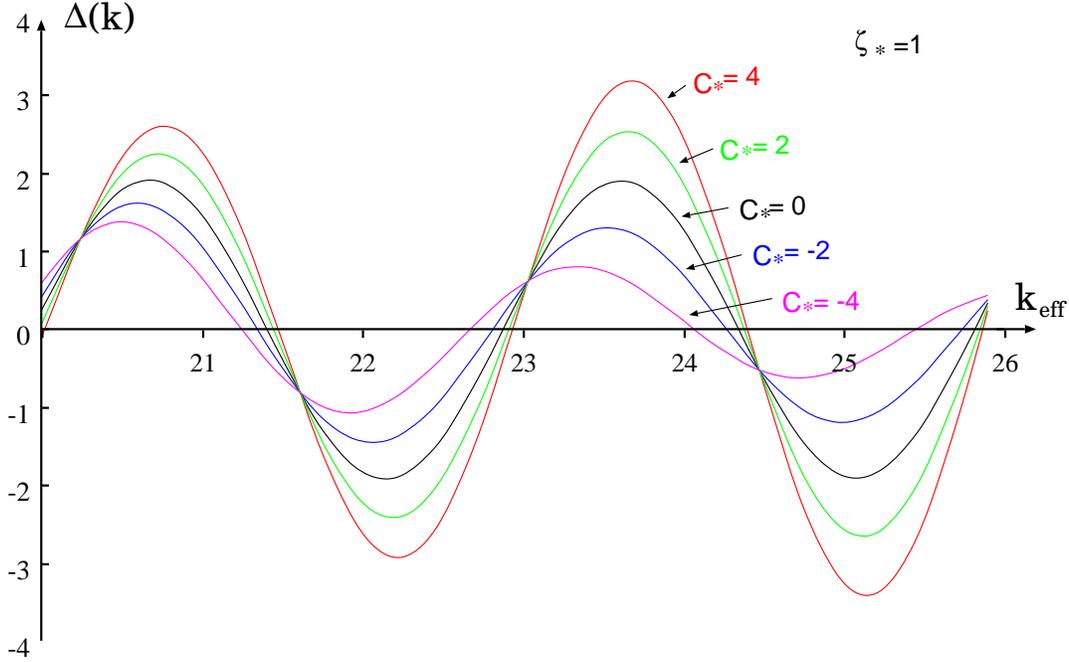}
\end{center}
  \caption{$\Delta(\k)$ at $\rho_r/\rho_m=0.1$ with $C_{\ast}=0
  , \: 2 \zeta_{\ast}, \: 4\zeta_{\ast}, 
\: -2 \zeta_{\ast},\:  -4 \zeta_{\ast}$ 
where we take $\zeta_{\ast}=1$. The horizontal coordinate
is the value $\k_{eff}=\dot{\alpha}_0^{-1} e^{-\alpha_0} \k$ 
at $\rho_r/\rho_m=1$. The 
perturbation with $\k_{eff}=1$ crosses the horizon at $\rho_r/\rho_m=1$. The initial conditions are set at $\rho_r/\rho_m=100$.}  
  \label{fig1}
\end{figure}

\newpage
\section{Conclusion}
\setcounter{equation}0
\hspace{0.5cm}
In this paper we obtained the effective 4D Einstein equations 
(\ref{3-6}) and (\ref{3-7}) 
that describe the scalar cosmological perturbations on the brane.
Then we investigated the effect of the bulk gravitational field
on the evolution of the cosmological perturbations on the brane.

We first used the equations on the brane obtained from the
power series expansion of the 5D Einstein equations.
From the equations on the brane, we obtained the effective Einstein
equations (\ref{2-9}), (\ref{2-10}) and (\ref{2-11}).
It should be mentioned that we cannot derive
the equation that corresponds to $(i \neq j)$ component of the 
4D Einstein equations which gives the relation between
the metric perturbations $\Phi_0$ and $\Psi_0$. 
Two types of the corrections are found.
One is written by the gradient of the metric perturbations. 
Another is independent of the metric perturbations 
(\ref{2-11}) and induces the density perturbations which behave 
as the sound waves with the sound velocity $1/\sqrt{3}$ 
at low energies. At large scales, they are 
homogeneous perturbations that depend only on time and decay like 
radiation. We identified them as the perturbations of the 
dark radiation. 

Then we derived the effective Einstein equations again 
in another way by solving the 
perturbations in the bulk and imposing the junction conditions 
(\ref{3-6}), (\ref{3-7}). We obtained the equation which gives the 
relation between $\Phi_0$ and $\Psi_0$.
We identified the perturbations in the
bulk which induce the perturbations of the dark radiation. 
These perturbations do not contribute to the metric perturbations but 
do contribute to the matter perturbations on the brane. 
At low energies, they have imaginary 
mass $2 \k^2 + 3 m^2=0$ in the bulk and diverge
at the horizon of the AdS spacetime.
The existence of them in the bulk depends on the boundary condition
of the perturbations.
We should impose two boundary conditions to completely determine 
the perturbations in the bulk. One is given by the equation of the 
state of the matter on the brane. The other choice of the boundary
condition at the horizon determines the existence of the perturbations 
of the dark radiation.

If we take the boundary condition that the perturbations
do not diverge at the horizon of the AdS spacetime, the 
perturbations of the dark radiation do not appear. 
The other corrections are suppressed by $\k l e^{-\alpha_0}$. 
Thus they correspond to the correction terms written by the
gradient of the metric perturbations. 
Corrections also arise in the relation between $\Phi_0$ and $\Psi_0$.
The corrections become large at the scales smaller than the
curvature scales of the AdS spacetime ($\k l e^{-\alpha_0} \gg 1$)
and in the high energy universe with the energy density 
larger than the tension of the brane ($\dot{\alpha}_0 l \gg 1$). 
Particularly, at high energies 
$\dot{\alpha}_0 l \gg 1$, the potential perturbation $\Phi_0$ 
becomes
dominant over the curvature perturbation $\Psi_0$.
We discuss the evolution of the adiabatic perturbations including 
these corrections. The interesting point is that at sufficiently 
large scales ($\k l e^{-\alpha_0} \to 0$)
the Bardeen parameter is constant even at high energies. 
Then the potential perturbation $\Phi_0$ are always constant 
if the barotropic parameter of the matter $w$ is constant. 
At the scales below $l$, 
the correction becomes large. In order to illustrate how these corrections 
modify the evolution of the density perturbations, we obtained the 
effective Poisson equation in the low energy universe at the sub-horizon scales. 
Using the effective Poisson equation, the evolution equation 
for the matter perturbations was given. The important point
is that the evolution equation becomes non-local once we
incorporate the effect of the perturbations in the bulk. 
This is the essential feature of the perturbations at scales
below $l$. We emphasized that one should be
careful to use the result $\k l e^{-\alpha_0} \to 0$ in the high energy
universe even at the super-horizon scales. This is because at high 
energies, the horizon scale of the universe itself is smaller than
the curvature scales $l$.

We should comment on our limitation in obtaining the
evolution of the perturbations using the effective Einstein
equations. It is in general difficult to obtain the spectrum
of the perturbations in the bulk $E^{(1)}(m,\k)$ 
by imposing the equation of state on the matter perturbations. 
As a result, we should make an assumption about the contribution of the
massive perturbations. We used the assumption
that the modes with $m l e^{-\alpha_0}>1 $ 
do not contribute to the perturbations in the bulk thus
take $m l e^{-\alpha_0} \to 0$. At low energies 
$e^{-\alpha_0} \to 0$, the assumption seems to be valid. 
$E^{(1)}(m,\k)$ is determined by the primordial
fluctuations and later evolutions \cite{KKJ}, \cite{SH}. 
Further studies are needed to 
know the exact form of $E^{(1)}(m,\k)$.

If we choose appropriate boundary conditions in the bulk (\ref{3-16-0}),
the perturbations of the dark radiation arise. They 
induce the isocurvature perturbations in the dust dominated universe. 
The key feature of them is that they can play 
a role even in the low energy universe at scales larger than $l$
where the former corrections are suppressed. 
We gave an evolution equation for the density perturbation 
including the corrections from them. The large scale CMB
anisotropies were estimated and the constraint on the amplitude
was derived. At sub-horizon scales, they act as an extra
force on the acoustic oscillations of the density perturbation.
In the matter dominated era, the location
and the height of the acoustic peak are shifted due to 
the extra force (see Fig.1). These shifts can be
directly observed by CMB anisotropies. Recently, many works have
been done about the test of the correlation between adiabatic 
and isocurvature perturbations using CMB spectrum \cite{Iso}. 
The detailed analysis of the CMB spectrum will reveal the 
existence of the dark sound waves.


\section*{Acknowledgements}
The work of K.K. was supported by JSPS Research Fellowships for
Young Scientist No.4687.

\appendix
\section{Background equations}
\setcounter{equation}0
In this Appendix, we derive the equations used in section 2.
The 5D Einstein equations are given by
\begin{equation}
G^M_N= \frac{6}{l^2} \delta^M_N + \kappa^2 
\frac{\sqrt{-g_{brane}}}{\sqrt{-g}} T^M_N 
=  \frac{6}{l^2} \delta^M_N 
+ e^{-\beta} \kappa^2 T^M_N  , 
\:\: (M,N =y,t,x^i). 
\label{A-1}
\end{equation}
We take the energy momentum tensor in the 5D spacetime as
\begin{equation}
T^M_N= \left(- \frac{6}{\kappa^2 l} diag(0,1,1,1,1)+
diag(0,-\rho,p,p,p) \right) \delta(y). 
\label{A-2}
\end{equation}
The Einstein tensor is given by
\begin{eqnarray}
G^0_{\: 0} &=& - 3 e^{-2 \beta}(\dot{\alpha}^2 + \dot{\alpha} \dot{\beta} 
-\alpha''-2 \alpha'^2 + \alpha' \beta') ,\nonumber\\
G^y_{\: y} &=& 3 e^{-2 \beta}
( -\ddot{\alpha}-2 \dot{\alpha}^2 + \dot{\alpha} \dot{\beta}
+ \alpha'^2 + \alpha' \beta') ,\nonumber\\
G^0_{\: y} &=&  -3 e^{-2 \beta}
( \beta' \dot{\alpha}+ \alpha' \dot{\beta} - \dot{\alpha}'
-\dot{\alpha} \alpha') ,\nonumber\\
G^{i}_{\:j} &=& 
\delta^{i}_{j} e^{-2 \beta}(-2 \ddot{\alpha} -3 \dot{\alpha}^2
- \ddot{\beta} + 2 \alpha''+ 3 \alpha'^2 + \beta'') .
\label{A-3}
\end{eqnarray}
In the $(0,0)$ and $(i,j)$ components of the 
Einstein equations, the jump of the first derivative 
of $\alpha(y,t)$ and 
$\beta(y,t)$ give the $\delta(y)$ function. 
These should be equated with the $\delta(y)$ function of the
matter. Then we obtain the junction conditions (\ref{1-4}) 
\begin{eqnarray}
\alpha_1(t) &=& - \frac{1}{l}   
- \frac{\kappa^2 \rho(t)}{6}, \nonumber\\
\beta_1(t) &=&  - \frac{1}{l}+ 
\frac{\kappa^2 \rho(t)}{3} + \frac{ \kappa^2 p(t)}{2},
\label{A-4}
\end{eqnarray}
where we take $e^{\beta_0}=1$.
The $y^0$-th order of the $(y,y)$ and $ (y,0)$ 
components of the Einstein equations give 
\begin{eqnarray}
-\ddot{\alpha}_0-2 \dot{\alpha}_0^2 +\alpha_1^2+\alpha_1 \beta_1
=\frac{2}{l^2}, \nonumber\\
\beta_1 \dot{\alpha}_0- \dot{\alpha}_1-\dot{\alpha}_0 \alpha_1
=0.
\end{eqnarray}
Using the junction condition (\ref{A-4}), 
we get (\ref{1-5}). 

Next let us derive the wave equations for $\beta$ and $\alpha$ 
in the bulk. The $(0,0)$ and $(i,j)$ components of 
the 5D Einstein equations in the bulk 
can be rewritten using $u=(t-y)/l$ and $v=(t+y)/l$ as
\begin{eqnarray}
\alpha_{,uv}+3 \alpha_{,u} \alpha_{,v}+e^{2 \beta}=0,
\nonumber\\
\beta_{,uv}-3 \alpha_{,u} \alpha_{, v}-\frac{1}{2}e^{2 \beta}=0.
\label{A-6}
\end{eqnarray}
We assume the bulk is AdS spacetime, so we take
$C_{M y N y}=0$ where $C_{MNKL}$ is the Weyl tensor. 
This condition is given by
\begin{equation}
\alpha_{,uv}-\beta_{,uv}=0.
\end{equation}
Then the wave equations for $\alpha$ and $\beta$ are 
given by
\begin{eqnarray}
\beta_{,uv}+\frac{1}{4} e^{2 \beta} &=& 0, \nonumber\\
\alpha_{,uv}+\frac{1}{4} e^{2 \beta}&=& 0,
\end{eqnarray}
which are (\ref{1-8}). Instead of solving the wave equations
directly, it is convenient to rewrite  
the equation for $\alpha$ using (\ref{A-6}) as 
\begin{equation}
\alpha_{,uv}-\alpha_{,u} \alpha_{,v}=0. 
\end{equation}
The solution can be found easily as
\begin{equation}
e^{\alpha}=\frac{1}{f(u)-g(v)},
\end{equation}
where $f(u)$ and $g(v)$ are the arbitrary functions.
Then $\beta$ can be obtained from (\ref{A-6}) as  
\begin{eqnarray}
e^{2 \beta}&=&-\alpha_{,uv}-3\alpha_{,u} \alpha_{,v}
\nonumber\\
&=& 4 \frac{f'(u)g'(v)}{(f(u)-g(v))^2}.
\end{eqnarray}

Finally, we show some background equations which are used in
the calculations of the perturbations. From the junction 
conditions (\ref{A-4}) and equations on the brane 
(\ref{1-5}), we can show the following equations;
\begin{eqnarray}
\ddot{\alpha}_0&=&\frac{\kappa^2 \alpha_1}{2} (\rho+p)
=\alpha_1 (\beta_1-\alpha_1), \nonumber\\
\alpha_1^2&=&\frac{1}{l^2}+\dot{\alpha}_0^2, \nonumber\\
\frac{\beta_1}{\alpha_1} &=& 1+\frac{\ddot{\alpha_0}}{\alpha_1^2},
\nonumber\\
\dot{\alpha}_1 &=&\frac{\dot{\alpha}_0 \ddot{\alpha}_0}{\alpha_1}
=\dot{\alpha}_0(\beta_1-\alpha_1).
\label{A-12}
\end{eqnarray}
In the calculations of the perturbations, we need $\alpha_2$ and
$\beta_2$. 
From the $y^0$-th order of $(0,0)$ and $(i,j)$ components of the 
Einstein equations, we can write 
$\alpha_2$ and $\beta_2$ in terms of $\alpha_0$, $\alpha_1$ 
and $\beta_1$;
\begin{eqnarray}
\alpha_2 &=& \dot{\alpha}_0^2 -2 \alpha_1^2 + \alpha_1 \beta_1 
+ \frac{2}{l^2},\nonumber\\
\beta_2  &=& \dot{\alpha}_0^2+2 \ddot{\alpha}_0+\alpha_1^2
-2 \alpha_1 \beta_1 + \frac{2}{l^2}.
\label{A-13}
\end{eqnarray}

\section{Derivation of the effective Einstein equation 
(\ref{3-6}) and (\ref{3-7})} 
\setcounter{equation}0
In this Appendix we review the formalism to solve the
perturbations in the bulk and impose the junction conditions
developed in \cite{KJ}. Using the formalism, we obtain the 
effective Einstein equations (\ref{3-6}) and (\ref{3-7}).

\subsection{Review of the formalism}
\hspace{0.5cm}
First let us review the formalism to obtain the perturbations
in the bulk. We start with the perturbed AdS spacetime
in Poincare coordinate; 
\begin{equation}
ds^2= \left(\frac{l}{z} \right)^2
\left(dz^2 - (1+2 \phi) d \tau^2 
+2 b_{,i} dx^i d \tau
+ \left((1 - 2 \hat{\Psi}) \delta_{ij}+ 
2 \hat{E}_{,ij} \right)dx^i dx^j\right), 
\end{equation}
Here $\phi,b,\hat{\Psi}$ and $\hat{E}$ is given by
\begin{eqnarray}
h = \left( \frac{z}{l}\right)^2 \int \frac{d^3 \k}{(2 \pi)^3}
\int d m \:\:
h (m,\k)  Z_2 (m z) e^{-i \omega \tau}e^{i \k \x}, 
(h= \phi,b,\hat{\Psi},\hat{E}),
\label{B-2}
\end{eqnarray}
where $Z_2$ is the combination of the Hankel function of the
first kind and the second kind of the second rank 
$Z_2(m z)=H^{(1)}_2(m z)+a(m) H^{(2)}_2(m z)$.
Here we used the transverse traceless gauge conditions
\begin{eqnarray}
&& \phi-3 \hat{\Psi} + \nabla^2 \hat{E} = 0, \nonumber\\
&& 2 \frac{d \phi}{d \tau} + \nabla^2 b =0, \nonumber\\
&& \frac{d b}{d \tau}+ 2 \hat{\Psi} -2 \nabla^2\hat{E}=0. 
\end{eqnarray}
Thus the coefficients $h (m,\k)$ satisfy 
\begin{eqnarray}
\phi (m,\k) &=& \frac{2 \k^4}{3 m^2} l^2  E(m,\k), \nonumber\\
b (m,\k) 
&=& -4 i \frac{\sqrt{\k^2+m^2} \:\: \k^2 l^2}{3 m^2} 
E (m,\k), \nonumber\\
\hat{\Psi}(m,\k) &=& 
-\frac{\k^2 l^2}{3}  E (m,\k), \nonumber\\
\hat{E} (m,\k)&=& \frac{2 \k^2 +3 m^2}{3 m^2} l^2 E(m,\k),
\label{B-4}
\end{eqnarray}
where $E(m,\k)$ is the arbitrary coefficient.
\hspace{0.5cm}\\
The perturbations in the metric (\ref{1-1}) 
is obtained by the coordinate transformation 
\begin{equation}
z=z(y,t)=l e^{- \alpha(y,t)},\quad \tau=\tau(y,t).
\end{equation}
The Jacobian of the transformation is given by
\begin{eqnarray}
\frac{\partial \tau}{\partial y} 
&=& l \dot{\alpha} e^{-\alpha},\:\:\:\:\:\:\:  
\frac{\partial z}{\partial y} =
-l \alpha' e^{-\alpha}, \nonumber\\
\frac{\partial \tau}{\partial t}
&=& l \alpha'  e^{-\alpha},\quad \:\: 
\frac{\partial z}{\partial t} 
= -l \dot{\alpha}  e^{-\alpha}.
\label{B-6}
\end{eqnarray}
Note that at late times 
\begin{equation}
\frac{dT}{d t}=l \alpha_1 e^{-\alpha_0}=
- e^{-\alpha_0}.
\label{B-7}
\end{equation}
Then $T=\tau(0,t)=- \eta$ where $\eta$ is the conformal time.
After the coordinate transformation, the resulting metric 
is given by 
\begin{eqnarray}
ds^2 &=&  e^{2 \beta(y,t)} \left((1+2 \hat{N}) dy^2 -(1+2 \hat{\Phi})
d t^2 +2 \hat{A} \: dt \: dy \right) \nonumber\\
&& +  e^{2 \alpha(y,t)}
\left(\left((1 - 2 \hat{\Psi}) \delta_{ij}+ 2 \hat{E}_{,ij}\right) 
dx^i dx^j+2 \hat{B}_{,i} dx^i dt + 2 \hat{G}_{,i} dx^i dy \right),
\label{B-8}
\end{eqnarray}
where
\begin{eqnarray}
\hat{\Phi} &=&  (l \alpha')^2 e^{-2 \beta} \phi, \nonumber\\
\hat{B} &=&  (l \alpha') e^{-\alpha} b ,\nonumber\\
\hat{N} &=& - (l \dot{\alpha})^2 e^{-2 \beta} \phi ,\nonumber\\
\hat{A} &=& -2 (l^2 \dot{\alpha} \alpha') e^{-2 \beta} \phi ,\nonumber\\
\hat{G} &=& (l \dot{\alpha}) e^{- \alpha} b.
\label{B-9}
\end{eqnarray}
In the metric (\ref{B-8}) the brane is not located at $y=0$. This
is because the matter perturbations on the brane bend the
brane. Then once we include the matter perturbations on the
brane the brane will be located at $y \neq 0$. The metric 
perturbations that we observe are those evaluated on the brane.
Then we should perform the (infinitesimal) coordinate 
transformation by
\begin{equation}
x^M \to x^M + \xi^M, \quad \xi^M=(\xi^y,\xi^t,\xi^{,i}),
\end{equation}
to ensure that the brane is located at $y=0$ in the new 
coordinate. 
We will take the gauge condition $G=A=0$ and $B_0=0,E_0=0$. 
This determines $\xi^t$ and $\xi$ in terms of $\xi^y$ as
\begin{eqnarray}
\xi^t &=& \int^y_0 dy (\hat{A}+ \dot{\xi}^y) + \hat{T}_0, \quad \quad 
\hat{T}_0 = e^{2 \alpha_0} (\hat{B}_0-\dot{\hat{E}}_0), \nonumber\\
\xi &=& -\int^y_0 dy (\hat{G}+e^{2(\beta-\alpha)} \xi^y) 
-\hat{E}_0.
\label{B-11}
\end{eqnarray}
Then we obtain the metric perturbations on the brane
\begin{eqnarray}
\Phi_0 &=& \hat{\Phi}_0+ 
\beta_1 \xi^y_0 +\dot{\hat{T}}_0, \nonumber\\
\Psi_0 &=& \hat{\Psi}_0- \alpha_1 \xi_0^y -\dot{\alpha}_0 \hat{T}_0, \nonumber\\
N_0 &=& \hat{N}_0 + \xi^{y}_1 + \beta_1 \xi^y_0, 
\label{B-12}
\end{eqnarray}
and the first derivative of the metric perturbations
\begin{eqnarray}
\Phi_1 &=& \ddot{\xi}^y_0 + \beta_1 \xi^{y}_1+ \beta_2 \xi_0^y 
+ \hat{\Phi}_1 + \dot{\hat{A}}_0+ \dot{\beta}_1 \hat{T}_0,
\nonumber\\
\Psi_1 &=& - \alpha_1 \xi^{y}_1-\dot{\alpha}_0 \dot{\xi}^y_0-\alpha_2 \xi^y_0
+ \hat{\Psi}_1 - \dot{\alpha}_0 \hat{A}_0 -\dot{\alpha}_1 \hat{T}_0
,\nonumber\\
N_1 &=&  \xi^{y}_2+ \beta_1 \xi^y_1 + \beta_2 \xi^y_0 +
\hat{N_1}+ \dot{\beta}_1 \hat{T}_0 
,\nonumber\\
B_1 &=&  e^{-2 \alpha_0}(-2 \dot{\xi}^y_0 +2 \dot{\alpha}_0 \xi^y_0
- 2 (\beta_1-\alpha_1) \hat{T}_0 
- \hat{A}_0 +e^{2 \alpha_0} \hat{B}_1 - e^{2 \alpha_0} \dot{\hat{G}}_0
) ,\nonumber\\
E_1 &=& \hat{E}_1 -e^{-2 \alpha_0} \xi^y_0 -\hat{G}_0.
\label{B-13}
\end{eqnarray}
Combining (\ref{B-13}) with the junction conditions \cite{KJ}
\begin{eqnarray}
\Psi_1 &=& -\alpha_1 N_0 +\frac{1}{6} \kappa^2 
\delta \rho, \nonumber\\
\Phi_1 &=&  \beta_1 N_0 + \kappa^2 
\left(\frac{\delta \rho}{3}+
\frac{\delta p}{2} \right), \nonumber\\
B_1 &=&  -2 (\beta_1-\alpha_1) e^{-\alpha_0} v, \nonumber\\
E_1 &=&  0,
\end{eqnarray}
we can write matter perturbations in terms of
$\xi^y_0$ and $E(m,\k)$;
\begin{eqnarray}
\kappa^2 \delta \rho &=& -6 \left( \dot{\alpha}_0 \dot{\xi}^y_0 
+(\alpha_2 -\alpha_1 \beta_1) \xi^y_0 
-\hat{\Psi}_1 + \dot{\alpha}_0 \hat{A}_0 + \dot{\alpha}_1 \hat{T}_0
-\alpha_1 \hat{N}_0 \right), \nonumber\\
\kappa^2 \delta p &=& 2 \left( \ddot{\xi}_0^y + 2 \dot{\alpha}_0 
\dot{\xi}_0^y 
+(2 \alpha_2 + \beta_2 -\beta_1^2 -2 \alpha_1 \beta_1)\xi_0^y 
\right. \nonumber\\
&& \left. + \hat{\Phi}_1 -2 \hat{\Psi}_1 + \dot{\hat{A}}_0
+2 \dot{\alpha}_0 \hat{A}_0 
+(\dot{\beta}_1+2 \dot{\alpha}_1) \hat{T}_0 
- (\beta_1+2 \alpha_1)\hat{N}_0 \right), \nonumber\\
\kappa^2(\rho+p) e^{\alpha_0} v &=& 2 \dot{\xi}_0^y -2 \dot{\alpha}_0 \xi^y_0 
- e^{2 \alpha_0} \hat{B}_1 
+ 2 (\beta_1-\alpha_1) \hat{T}_0 +e^{2 \alpha_0} \dot{\hat{G}}_0 + \hat{A}_0 
,\nonumber\\
0&=& -2 e^{-2 \alpha_0} \xi^y_0 +2 \hat{E}_1 -2 \hat{G}_0,
\label{B-15}
\end{eqnarray}
where $\hat{T}_0= e^{2 \alpha_0} (\hat{B}_0-\dot{\hat{E}}_0)$.
From the last equation in (\ref{B-15}), $\xi^y_0$ is written by
$E(m,\k)$. Thus the matter perturbations are written in terms of the 
perturbations in the bulk (\ref{B-2}), (\ref{B-4}). 
These equations correspond to (\ref{1-10}) in the background 
spacetime. 
The solutions of the perturbations are obtained by determining 
$E(m,\k)$ and $a(m)$ by imposing the equations of state of 
the matter 
perturbations such as $\delta p = c_s^2 \delta \rho$
and the appropriate boundary condition in the bulk.
In \cite{KJ}, $E(m,\k)$ is obtained for perturbations 
at the super-horizon scales in the low energy universe with 
the constant barotropic parameter with the boundary condition that
the perturbations are out-going at the horizon of the 
AdS spacetime. 
In general, however, it is rather difficult 
to obtain the solution for $E(m,\k)$. 
Thus we use the method described
in section 2. We rewrite (\ref{B-15}) 
into the effective Einstein equations. To do so,
we should rewrite the right-hand side of (\ref{B-15}) in terms of 
the metric perturbations $\Phi_0$ and $\Psi_0$. 

\subsection{Derivation of the equations 
(\ref{3-6}) and (\ref{3-7})}
\hspace{0.5cm}
We rewrite the right-hand
side of the equations (\ref{B-15}) by the metric perturbations
$\Phi_0$ and $\Psi_0$ to derive 
(\ref{3-6}) and (\ref{3-7}). 
We will write $\hat{\Phi}$, $\hat{B}$, $\hat{N}$, $\hat{A}$ and $\hat{G}$
by $\phi$ and $b$ using (\ref{B-9}). 
First let us consider the density perturbation $\delta \rho$. 
From (\ref{B-15}), $\delta \rho$ is given by
\begin{equation}
\kappa^2\delta \rho=-6 \left(\dot{\alpha_0} \dot{\xi}^y_0
-\dot{\alpha}_0^2
\xi^y_0-\dot{\alpha}_0^2 \alpha_1 l^2 \phi_0
+\dot{\alpha_0} \ddot{\alpha}_0 e^{\alpha_0} l b_0-
\frac{\dot{\alpha}_0 \ddot{\alpha}_0}{\alpha_1} e^{2 \alpha_0}
\dot{\hat{E}}_0 -\hat{\Psi}_1\right),
\label{B-16}
\end{equation}
where we used (\ref{A-13}) to write $\alpha_2-\alpha_1 \beta_1=-
\dot{\alpha_0 }^2$ and $\dot{\alpha_1}=\dot{\alpha_0}
\ddot{\alpha}_0/\alpha_1$. 
The strategy is
to write $\xi^y_0$ by $\Psi_0$ and $\Phi_0$.
From (\ref{B-12}), the metric perturbations
$\Psi_0$ and $\Phi_0$ are given by
\begin{eqnarray}
\Psi_0 &=& \hat{\Psi}_0-\alpha_1 \xi_0^y
+\dot{\alpha}_0 e^{2 \alpha_0} \dot{\hat{E}}_0
-\dot{\alpha}_0 \alpha_1 e^{\alpha_0}b_0, \nonumber\\
\Phi_0 &=& \alpha_1^2 l^2\phi_0 
+\left(1+\frac{\ddot{\alpha}_0}{\alpha_1^2} \right)\alpha_1
\xi_0^y
-e^{2 \alpha_0} \ddot{\hat{E}}_0- 2 \dot{\alpha}_0
e^{2 \alpha_0} \dot{\hat{E}}_0+(\alpha_1 \dot{\alpha}_0
+\dot{\alpha}_1)l e^{\alpha_0} b_0 +\alpha_1 e^{\alpha_0}l
\dot{b}_0, \nonumber\\
\label{B-17}
\end{eqnarray}
where we used (\ref{A-12}) to write $\beta_1=(1+\ddot{\alpha}_0/
\alpha_1^2) \alpha_1$. 
From (\ref{B-17}), we can show 
\begin{equation}
\dot{\alpha}_0 \dot{\Psi}_0+\dot{\alpha}_0^2
\Phi_0 =-\alpha_1 \left(\dot{\alpha}_0 \dot{\xi}^y_0
-\dot{\alpha}_0^2 \xi^y_0-
\dot{\alpha}_0^2 \alpha_1 l^2 \phi_0
+\dot{\alpha}_0 \ddot{\alpha}_0 e^{\alpha_0} l b_0
-\frac{\dot{\alpha}_0 \ddot{\alpha}_0}{\alpha_1}
 e^{2 \alpha_0} \dot{\hat{E}}_0 \right)
+\dot{\alpha}_0 \dot{\hat{\Psi}}_0.
\end{equation}
Then the terms written by $\xi^y_0$ in
(\ref{B-16}) can be rewritten by $\Phi_0$ and $\Psi_0$.  
We obtain 
\begin{equation}
-\frac{\alpha_1 \kappa^2}{2}\delta \rho =-3 \left(
\dot{\alpha}_0 \dot{\Psi}_0+\dot{\alpha}_0^2 \Phi_0
+ \alpha_1 \hat{\Psi}_1
-\dot{\alpha_0} \dot{\hat{\Psi}}_0 \right).
\label{B-23}
\end{equation}
The remaining task is to rewrite the terms written by $\hat{\Psi}$
in terms of $\Psi_0$. First, using the solution of the perturbations
(\ref{B-2}) and (\ref{B-4}), we rewrite $\Psi_0$ in terms of
$E(m,\k)$. From (\ref{B-15}), $\xi^y_0$ is given by
\begin{equation}
\xi^y_0=e^{2 \alpha_0} \hat{E}_1- \dot{\alpha}_0
e^{\alpha_0} l b_0.
\label{B-23-1}
\end{equation}
Then we can rewrite $\Psi_0$ as
\begin{equation}
\Psi_0 
=\hat{\Psi}_0-\alpha_1 e^{2 \alpha_0}\hat{E}_1
+\dot{\alpha}_0 e^{2 \alpha_0} \dot{\hat{E}}_0. 
\end{equation}
Using the Jacobian of the transformation (\ref{B-6}) and 
\begin{equation}
\frac{d}{dz}(z^2 Z_2(mz))= m z^2 Z_1(mz), 
\end{equation}
we can get the following equations;
\begin{eqnarray}
\alpha_1 e^{2 \alpha_0} \hat{E}_1(\k)
&=& -\int dm \hat{E}(m,\k)
 \left(\alpha_1^2 m l e^{-\alpha_0}
Z_1( mle^{-\alpha_0})+\dot{\alpha}_0 \alpha_1 
i \omega l e^{-\alpha_0}
Z_2(m l e^{-\alpha_0}) \right) 
e^{-i \omega T}, \nonumber\\
\dot{\alpha}_0 e^{2 \alpha_0} \dot{\hat{E}}_0(\k)
&=& -\int dm  \hat{E}(m,\k)
\left(\dot{\alpha_0}^2 m l e^{-\alpha_0}
 Z_1( mle^{-\alpha_0})+\dot{\alpha}_0 \alpha_1 
i \omega l e^{-\alpha_0}
Z_2(m l e^{-\alpha_0}) \right) 
e^{-i \omega T}. \nonumber\\
\label{B-26}
\end{eqnarray}
Then we find
\begin{equation}
-\alpha_1 e^{2 \alpha_0}\hat{E}_1
+\dot{\alpha}_0 e^{2 \alpha_0} \dot{\hat{E}}_0
=\int dm E(m,\k)\left(\frac{2 \k^2 +3 m^2}{3 m}l e^{-\alpha_0}
\right) Z_1(m l e^{-\alpha_0})e^{- i \omega T},
\label{B-27}
\end{equation}
where we used (\ref{B-4}) and 
$\alpha_1^2=1/l^2+\dot{\alpha}_0^2$ (\ref{A-12}).
$\hat{\Psi}$ is given by
\begin{eqnarray}
\hat{\Psi}(\k) &=& e^{-2 \alpha_0}\int dm \hat{\Psi}(m,\k) 
Z_2(m l e^{-\alpha_0}) e^{-i \omega T} \nonumber\\
&=& -\int dm  E(m,\k) \left(
\frac{2 \k^2}{3 m}l e^{-\alpha_0} Z_1(m l e^{-\alpha_0})
-\frac{1}{3}
(\k l e^{-\alpha_0})^2 Z_0(m l e^{-\alpha_0}) \right)
e^{- i \omega T},
\end{eqnarray}
where we used (\ref{B-4}) and $Z_2(mz)=(2/mz)Z_1(mz)-Z_0(mz)$.
Then we can write $\Psi_0$ in terms of $E(m,\k)$ as
\begin{equation}
\Psi_0(\k)= \int dm E(m,\k)\left(
m l e^{-\alpha_0} Z_1 (m l e^{-\alpha_0})
+ \frac{1}{3}  (\k l e^{-\alpha_0})^2 Z_0(m l 
e^{-\alpha_0})\right) e^{-i \omega T(t)}.
\label{B-29}
\end{equation}
On the other hand, the same calculations as (\ref{B-27}) 
yields
\begin{equation}
\alpha_1 \hat{\Psi}_1(\k) -\dot{\alpha}_0 \dot{\hat{\Psi}}_0(\k)
= \frac{1}{3}\int dm E(m,\k) \k^2 m l e^{-3 \alpha_0}
Z_1(m l e^{-\alpha_0}) e^{-i \omega T(t)}.
\end{equation}
Then (\ref{B-23}) becomes
\begin{eqnarray}
-\frac{\kappa^2 \alpha_1}{2} \delta \rho (\k)
&=& -3 (\dot{\alpha}_0 \dot{\Psi}_0 + \dot{\alpha}_0^2 \Phi_0)
-e^{-2 \alpha_0} \k^2 \Psi_0 \nonumber\\
&+&  
\frac{1}{3}e^{-4 \alpha_0} \int dm E(m,\k) \k^4 l^2
Z_0(m l e^{-\alpha_0}) e^{-i \omega T(t)}.
\label{B-31}
\end{eqnarray}

The other quantities $\delta p$ and $v$ 
can be calculated in the same way. 
The calculations are straightforward but lengthy.
It is easier to derive $\delta p$ and $v$ using the equations 
on the brane (\ref{2-3}), (\ref{2-4}) and (\ref{2-5}). 
The pressure perturbations $\delta p$ can be obtained 
from (\ref{2-3}) as
\begin{equation}
-\frac{\kappa^2 \alpha_1}{2} \delta p (\k)
= \ddot{\Psi}_0+4 \dot{\alpha}_0 \dot{\Psi}_0 +\dot{\alpha}_0 \dot{\Phi}_0
+2(\ddot{\alpha}_0+2 \dot{\alpha}_0^2) \Phi_0 
+\frac{1}{3} e^{-2 \alpha_0} (2 \k^2 \Psi_0 -\k^2 \Phi_0)
-\frac{\kappa^2 \beta_1}{6} \delta \rho(\k).
\label{B-32}
\end{equation}
Substituting (\ref{B-31}) into (\ref{B-32}), 
we get $\delta p$ as
\begin{eqnarray}
-\frac{\kappa^2 \alpha_1}{2} \delta p(\k)
&=& \ddot{\Psi}_0+\left(3 \dot{\alpha}_0-\frac{\dot{\alpha_0} 
\ddot{\alpha}_0}{\alpha_1^2}\right) \dot{\Psi}_0+
\dot{\alpha}_0 \dot{\Phi}_0 +\left( 2 \ddot{\alpha}_0
-\frac{\dot{\alpha}_0^2 \ddot{\alpha}_0}{\alpha_1^2} +3 \dot{\alpha}_0^2
\right) \Phi_0 \nonumber\\
&-& \frac{1}{3}e^{-2 \alpha_0}\k^2 \Phi_0+\frac{1}{3}
\left(1-\frac{\ddot{\alpha}_0}{\alpha_1^2} \right)
e^{-2 \alpha_0}\k^2 \Psi_0 \nonumber\\
&+& \frac{1}{9}
\left(1+\frac{\ddot{\alpha}_0}{\alpha_1^2} \right)
e^{-4 \alpha_0} \int dm E(m,\k) \k^4 l^2
Z_0(m l e^{-\alpha_0}) e^{-i \omega T(t)}.
\label{B-33}
\end{eqnarray}
The velocity perturbation $v$ is also obtained from the 
equation on the brane (\ref{2-5})
\begin{equation}
(\rho+p)e^{\alpha_0}v =
\k^{-2} e^{2 \alpha_0} \left(
-\dot{\delta \rho}+ 3 (\rho+p) \dot{\Psi}_0 
- 3\dot{\alpha}_0 (\delta \rho+\delta p) \right).
\label{B-34}
\end{equation}
Substituting (\ref{B-31}) and (\ref{B-33}) into (\ref{B-34}), 
we can show $v$ is given by
\begin{equation}
-\frac{\kappa^2 \alpha_1}{2} (\rho+p) e^{\alpha_0}v(\k) =
\dot{\Psi}_0 +\dot{\alpha}_0 \Phi_0 +  \Delta v,
\label{B-35}
\end{equation}
where $\Delta v$ is given by 
\begin{eqnarray}
\Delta v &=& -\k^{-2} e^{-2 \alpha_0}
\frac{d}{dt} \left[ \frac{1}{3}
\int dm E(m,\k) \k^4 l^2
Z_0(m l e^{-\alpha_0}) e^{-i \omega T(t)} \right] \nonumber\\
&=& \frac{1}{3} e^{-3\alpha_0} \int dm E(m,\k) \left( 
\alpha_1 i \omega \k^2 l^3 Z_0 
(m l e^{-\alpha_0}) -\dot{\alpha}_0 
m \k^2 l^3  Z_1(m l e^{-\alpha_0})
\right) e^{-i \omega T(t)}. \nonumber\\
\label{B-36}
\end{eqnarray}

Finally, let us rewrite 
the metric perturbations $\Phi_0$ in terms of 
$E(m,\k)$ to derive (\ref{3-7}). 
From (\ref{B-17}) and (\ref{B-23-1}), 
$\Phi_0$ is given by
\begin{equation}
\Phi_0=\alpha_1^2 l^2 \phi_0 
+\alpha_1 e^{\alpha_0} l \dot{b}_0
+ \left(1+
\frac{\ddot{\alpha}_0}{\alpha_1^2}\right) \alpha_1 e^{2 \alpha_0}
\hat{E}_1 -2  \dot{\alpha}_0 e^{2 \alpha_0} \dot{\hat{E}}_0
-e^{2 \alpha_0} \ddot{\hat{E}}_0.
\end{equation}
From (\ref{B-26}), 
\begin{eqnarray}
\lefteqn{\left(1+ \frac{\ddot{\alpha}_0}{\alpha_1^2}\right) 
\alpha_1 e^{2 \alpha_0}
\hat{E}_1
-2  \dot{\alpha}_0 e^{2 \alpha_0} \dot{\hat{E}}_0
-e^{2 \alpha_0} \ddot{\hat{E}}_0} \nonumber\\
&&= \int dm E(m,\k) \left(
(\alpha_1 l)^2
\frac{2\k^2+m^2}{m} l e^{-\alpha_0}
Z_1(m l e^{-\alpha_0}) 
- (\omega^2 \alpha_1^2+m^2 \dot{\alpha}_0^2)
l^4 e^{-2 \alpha_0}Z_0(m le^{-\alpha_0}) \right. \nonumber\\
&& \left. \quad \quad -2 (\alpha_1 \dot{\alpha}_0 l^2) i \omega m l^2
e^{-2 \alpha_0} Z_1(m l e^{-\alpha_0}) \right)
\frac{2\k^2+3 m^2}{3 m^2} e^{-i \omega T},
\end{eqnarray}
where we used $d (z Z_1(mz))/dz= m z Z_0(mz)$.
In addition, we can show
\begin{equation}
\alpha_1^2 l^2 \phi_0=(\alpha_1 l)^2
\int dm E(m,\k) \left(\frac{4 \k^4}{3 m^3}l
e^{-\alpha_0} Z_1(m l e^{-\alpha_0})
-\frac{2 \k^4}{3 m^2}l^2 e^{-2\alpha_0}
Z_0(m l e^{-\alpha_0}) \right) e^{-i \omega T},
\end{equation}
\begin{eqnarray}
\alpha_1 e^{\alpha_0} l \dot{b}_0 &=&
\int dm E(m,\k) \left(-(\alpha_1 l)^2
\frac{8 \omega^2 \k^2}{3 m^3}l e^{-\alpha_0}
Z_1(m l e^{-\alpha_0}) \right.\nonumber\\
&& \left. +(\alpha_1 l)^2 \frac{4 \omega^2 \k^2}{3 m^2}l^2
e^{-2\alpha_0} Z_0(m l e^{-\alpha_0})
+(\alpha_1 \dot{\alpha}_0 l^2) \frac{4 i \omega \k^2}{3 m}
l^2 e^{-2 \alpha_0} Z_1(m l e^{-\alpha_0}) \right)
e^{-i \omega T},\nonumber\\
\end{eqnarray}
where we used $Z_2(mz)=(2/mz)Z_1(mz)-Z_0(mz)$.
Then $\Phi_0$ can be written by $E(m,\k)$ as
\begin{eqnarray}
\Phi_0(\k)&=& 
\int dm E(m,\k) 
\left(
m l e^{-\alpha_0} Z_1 (m l e^{-\alpha_0})
-\frac{1}{3}(\k^2+3 m^2)l^2 e^{-2 \alpha_0}
Z_0(m l e^{-\alpha_0}) \right) e^{-i \omega T(t)}
\nonumber\\
&+&(\dot{\alpha}_0 l)^2 \int dm E(m,\k)
\left(mle^{-\alpha_0} Z_1(m l e^{-\alpha_0}) 
- (\k^2+2 m^2)l^2  e^{-2 \alpha_0}
Z_0(m l e^{-\alpha_0}) \right)e^{-i \omega T(t)} \nonumber\\
&-& 2 \alpha_1 \dot{\alpha}_0 l^2 \int dm E(m,\k) 
(i \omega m l^2 e^{-2 \alpha_0}) Z_1(m l e^{-\alpha_0}) 
e^{-i \omega T(t)},
\end{eqnarray}
where we used $(\alpha_1 l)^2=1+(\dot{\alpha}_0 l)^2$.

\section{Primordial fluctuations}
\setcounter{equation}0
\hspace{0.5cm}
The CMB anisotropies at large scales are determined  
by $\zeta_{\ast}$ which should be determined
by the primordial fluctuations. We consider the inflaton
$\phi$ confined to the brane with potential $V(\phi)$ \cite{In}.
The background equations are given by
\begin{eqnarray}
\ddot{\phi} + 3 \dot{\alpha}_0 \dot{\phi} &=& -\frac{d V(\phi)}{d \phi},
\nonumber\\
\frac{\kappa^2 \alpha_1}{2} \dot{\phi}^2
&=& \ddot{\alpha}_0.
\end{eqnarray}
The perturbed energy-momentum 
tensor of the inflaton is given by
\begin{eqnarray}
\delta \rho&=&-\dot{\phi}^2 \Phi_0+\dot{\phi}\dot{\delta \phi}
+V'(\phi)\delta \phi ,\nonumber\\
\delta p&=&-\dot{\phi}^2 \Phi_0+\dot{\phi}\dot{\delta \phi}
-V'(\phi)\delta \phi  ,\nonumber\\
(\rho+p)e^{\alpha_0}v &=& \dot{\phi} \delta \phi,
\label{C-1}
\end{eqnarray}
where $\delta \phi$ is the fluctuations of the inflaton.
It is useful to use the Mukhanov variable to describe
the evolution of the perturbations;
\begin{equation}
Q=\delta \phi +\frac{\dot{\phi}}{\dot{\alpha}_0} \Psi_0.
\end{equation}
Combining the equations (\ref{C-1}) and (\ref{3-6})
and equation of motion for $\delta \phi$
\begin{equation}
\ddot{\delta \phi}+3 \dot{\alpha}_0 \dot{\delta \phi}
+e^{-2 \alpha_0} \k^2 \delta \phi + V''(\phi)\delta \phi
=3 \dot{\phi} \dot{\Psi}_0+\dot{\phi} \dot{\Phi}_0-2 V'(\phi) \Phi_0,
\end{equation}
we can obtain the evolution equation for $Q$. 
We get
\begin{equation}
\ddot{Q} + 3 \dot{\alpha}_0 \dot{Q}+e^{-2 \alpha_0} \k^2 Q
+\left(\frac{\dot{\ddot{\alpha}}_0}{\dot{\alpha}_0}-
2 \frac{\ddot{\alpha}_0}{\dot{\alpha}_0}\frac{V'(\phi)}
{\dot{\phi}}-2 \left(\frac{\ddot{\alpha}_0}{\dot{\alpha}_0}
\right)^2+V''(\phi) \right)Q = J,
\label{C-4}
\end{equation}
where 
\begin{eqnarray}
J &=& \frac{\dot{\phi}_0}{\dot{\alpha}_0}
\left( \left(\frac{2\dot{\ddot{\alpha}}_0}{\ddot{\alpha}_0}-
\frac{\ddot{\alpha}_0}{\dot{\alpha}_0} \right) \Delta v 
+\frac{1}{3}e^{-2 \alpha_0}\k^2 \Phi_0 - \frac{1}{3}
\left(1-\frac{\ddot{\alpha}_0}{\alpha_1^2} \right)
e^{-2 \alpha_0}\k^2 \Psi_0 \right. \nonumber\\
&+& \left. \frac{1}{3} 
\left(\frac{2}{3}-\frac{1}{3} \frac{\ddot{\alpha}_0}{\alpha_1^2}
\right)e^{-4 \alpha_0} \int dm E(m,\k) \k^4 l^2
Z_0(m l e^{-\alpha_0}) e^{-i \omega T(t)} \right),
\end{eqnarray}
where $\Delta v$ is given by (\ref{B-36}). 
We take the boundary condition so that 
$\delta \rho_{\chi}=\delta p_{\chi}=v_{\chi}=0$.
At large scales $J \to 0$, then we can find the solution for $Q$ as
\begin{equation}
Q=\frac{\dot{\phi}}{\dot{\alpha}_0} \left(A_{Q}
+B_{Q} \int^t dt' \frac{\dot{\alpha_0}^2}{e^{3 \alpha_0}
\dot{\phi}^2} \right),
\end{equation}
where $A_{Q}$ and $B_{Q}$ is the constant of the integration.
The amplitude of the growing mode solution $A_{Q}$ 
is determined once $Q$ is quantised. Denoting the
power spectrum of $A_{Q}$ as $P_{A_{Q}}$, we get
\begin{equation}
P_{A_{Q}}=\frac{\dot{\alpha}_0}{\dot{\phi}} \left. 
P_Q \right \vert_{large \:scales},
\end{equation}
where right-hand side is the power spectrum of the
quantised $Q$ evaluated at large scales. 
The important point is that $Q$ is related to the
Bardeen parameter by (\ref{C-1}), (\ref{3-6}) and 
(\ref{4-3}) as
\begin{equation}
Q=\frac{\dot{\phi}}{\dot{\alpha}_0}\left(
\zeta-
\frac{\dot{\alpha}_0}{\ddot{\alpha}_0}
\Delta v \right).
\end{equation}
Then at large scales $A_{Q}=\zeta_{\ast}$ and 
\begin{equation}
P_{\zeta_{\ast}}=\left. P_Q \right \vert_{large\: scales}.
\end{equation}
The problem is how to quantise the system of (\ref{C-4}).
As in the evolution equation for the density perturbations, 
the equation becomes non-local at scales below $l$
($\k l e^{-\alpha_0} >0$). Particularly, at high energies, 
the Hubble horizon is smaller than the curvature scale $l$.
Thus even at the horizon scale, the corrections are significant.
One way is to construct the effective action which gives the 
equation (\ref{C-4}) and do path-integral quantisation 
as is done in \cite{Haw}. Further investigations are 
needed to obtain the spectrum of $\zeta_{\ast}$.

\end{document}